\documentclass[english]{article}
\usepackage[T1]{fontenc}
\usepackage[utf8]{inputenc}

\usepackage[style=ieee, doi=false, isbn=false, url=false]{biblatex}
\usepackage{hyperref}
\usepackage{bm,bbm}
\usepackage{amsthm,amssymb,amsmath,mathtools}
\usepackage{stmaryrd}
\usepackage{microtype}
\usepackage{algpseudocode,algorithm,algorithmicx}
\usepackage{graphicx,tabularx,float}
\usepackage[svgnames]{xcolor}
\usepackage{babel,csquotes}
\usepackage{comment}

\makeatletter

\theoremstyle{plain}
\newtheorem{theorem}{Theorem}

\newtheorem{lemma}[theorem]{Lemma}

\newtheorem*{remark}{Remark}

\bmdefine{\bA}{A}
\bmdefine{\ba}{a}
\bmdefine{\bB}{B}
\bmdefine{\bb}{b}
\bmdefine{\bC}{C}
\bmdefine{\bc}{c}
\bmdefine{\bD}{D}
\bmdefine{\bd}{d}
\bmdefine{\bE}{E}
\bmdefine{\be}{e}
\bmdefine{\bF}{F}
\bmdefine{\bG}{G}
\bmdefine{\bg}{g}
\bmdefine{\bH}{H}
\bmdefine{\bh}{h}
\bmdefine{\bI}{I}
\bmdefine{\bi}{i}
\bmdefine{\bJ}{J}
\bmdefine{\bj}{j}
\bmdefine{\bK}{K}
\bmdefine{\bk}{k}
\bmdefine{\bL}{L}
\bmdefine{\bl}{l}
\bmdefine{\bM}{M}
\bmdefine{\bN}{N}
\bmdefine{\bn}{n}
\bmdefine{\bO}{O}
\bmdefine{\bo}{o}
\bmdefine{\bP}{P}
\bmdefine{\bp}{p}
\bmdefine{\bQ}{Q}
\bmdefine{\bq}{q}
\bmdefine{\bR}{R}
\bmdefine{\br}{r}
\bmdefine{\bS}{S}
\bmdefine{\bs}{s}
\bmdefine{\bT}{T}
\bmdefine{\bt}{t}
\bmdefine{\bU}{U}
\bmdefine{\bu}{u}
\bmdefine{\bV}{V}
\bmdefine{\bv}{v}
\bmdefine{\bW}{W}
\bmdefine{\bw}{w}
\bmdefine{\bX}{X}
\bmdefine{\bx}{x}
\bmdefine{\bY}{Y}
\bmdefine{\by}{y}
\bmdefine{\bZ}{Z}
\bmdefine{\bz}{z}

\bmdefine{\balpha}{\alpha}
\bmdefine{\bbeta}{\beta}
\bmdefine{\bgamma}{\gamma}
\bmdefine{\bGamma}{\Gamma}
\bmdefine{\bdelta}{\delta}
\bmdefine{\bDelta}{\Delta}
\bmdefine{\bepsilon}{\epsilon}
\bmdefine{\bvarepsilon}{\varepsilon}
\bmdefine{\bzeta}{\zeta}
\bmdefine{\bmeta}{\eta}
\bmdefine{\btheta}{\theta}
\bmdefine{\bTheta}{\Theta}
\bmdefine{\biota}{\iota}
\bmdefine{\bkappa}{\kappa}
\bmdefine{\blambda}{\lambda}
\bmdefine{\bLambda}{\Lambda}
\bmdefine{\bmu}{\mu}
\bmdefine{\bnu}{\nu}
\bmdefine{\bpi}{\pi}
\bmdefine{\bPi}{\Pi}
\bmdefine{\brho}{\rho}
\bmdefine{\bsigma}{\sigma}
\bmdefine{\bSigma}{\Sigma}
\bmdefine{\btau}{\tau}
\bmdefine{\bupsilon}{\upsilon}
\bmdefine{\bUpsilon}{\Upsilon}
\bmdefine{\bphi}{\phi}
\bmdefine{\bPhi}{\Phi}
\bmdefine{\bchi}{\chi}
\bmdefine{\bpsi}{\psi}
\bmdefine{\bPsi}{\Psi}
\bmdefine{\bomega}{\omega}
\bmdefine{\bOmage}{\Omega}

\newcommand{\cD}{{\mathcal{D}}}

\newcommand{\cF}{{\mathcal{F}}}
\newcommand{\cI}{\mathcal{I}}

\newcommand{\cM}{\mathcal{M}}

\newcommand{\cS}{{\mathcal{S}}}

\newcommand{\cL}{\mathcal{L}}

\newcommand{\bbT}{\mathbb{T}}
\newcommand{\bbR}{\mathbb{R}}
\newcommand{\bbC}{\mathbb{C}}
\newcommand{\bbZ}{\mathbb{Z}}
\newcommand{\bbN}{\mathbb{N}}

\newcommand{\sT}{{\mathsf{T}}}

\newcommand{\sH}{{\mathsf{H}}}

\newcommand{\rd}{\mathrm{d}}

\DeclareMathOperator*{\argmin}{arg\,min}

\DeclareMathOperator{\diag}{diag}

\newcommand{\herm}{{\mathsf{H}}}

\newcommand{\pprime}{{\prime\prime}}
\newcommand{\ppprime}{{\prime\prime\prime}}

\newcommand\norm[1]{\left\lVert #1 \right\rVert}

\makeatother

\usepackage[margin=1in]{geometry}

\allowdisplaybreaks[1]

\hypersetup{unicode = true,
colorlinks,
linkcolor=blue,anchorcolor=blue,citecolor=blue,
pdftitle = {Local Geometry of Nonconvex Spike Deconvolution from Low-Pass Measurements},
pdfauthor = {Maxime Ferreira Da Costa, Yuejie Chi}
}

\title{Local Geometry of Nonconvex Spike Deconvolution \\ from Low-Pass Measurements}

\author{
	Maxime Ferreira Da Costa\thanks{M. Ferreira Da Costa is with the Laboratory of Signals and Systems (L2S) at CentraleSupélec, Université Paris-Saclay. Part of this work was done while he was with University of Southern California and Carnegie Mellon University. Email: \texttt{maxime.ferreira@centralesupelec.fr}.} \\
	CentraleSupélec | Université Paris-Saclay \\
	 \and
	Yuejie Chi\thanks{Y. Chi is with Department of Electrical and Computer Engineering, Carnegie Mellon University. The work of Y.~Chi is supported in part by Office of Naval Research under N00014-19-1-2404, and by National Science Foundation under CAREER ECCS-1818571 and ECCS-2126634. Email: \texttt{yuejiechi@cmu.edu}.} \\
	Carnegie Mellon University
}

\date{August 2022; Revised February 2023}

\begin{document}

\maketitle

\begin{abstract}
	Spike deconvolution is the problem of recovering the point sources from their convolution with a known point spread function, which plays a fundamental role in many sensing and imaging applications. In this paper, we investigate the local geometry of recovering the parameters of point sources---including both amplitudes and locations---by minimizing a natural nonconvex least-squares loss function measuring the observation residuals. We propose preconditioned variants of gradient descent (GD), where the search direction is scaled via some carefully designed preconditioning matrices. We begin with a simple fixed preconditioner design, which adjusts the learning rates of the locations at a different scale from those of the amplitudes, and show it achieves a linear rate of convergence---in terms of {\em entrywise} errors---when initialized close to the ground truth, as long as the separation between the true spikes is sufficiently large. However, the convergence rate slows down significantly when the dynamic range of the source amplitudes is large. To bridge this issue, we introduce an adaptive preconditioner design, which compensates for the learning rates of different sources in an iteration-varying manner based on the current estimate. The adaptive design provably leads to an accelerated convergence rate that is independent of the dynamic range, highlighting the benefit of adaptive preconditioning in nonconvex spike deconvolution. Numerical experiments are provided to corroborate the theoretical findings.
\end{abstract}


\tableofcontents

\section{Introduction}

Spike deconvolution, also known as super-resolution~\cite{donoho_superresolution_1992}, is the task of recovering a stream of point sources from their convolution with a point spread function (PSF). This classical problem is at the core of many sensing and imaging modalities, including but not limited to radar, sonar, optical imaging, neuroimaging, and communication systems \cite{potter_sparsity_2010,zhu_super-resolution_2016,zhu_faster_2012,berger_sparse_2010}. The PSF, which models the physical limitations of the imaging device involved in the experimental process, is commonly assumed to act as a band-limited and shift-invariant low-pass filter on the point sources~\cite{lindberg_mathematical_2012,born2013principles}. The sharpness of the original sources, modeled by Dirac impulses, is degraded through the convolution process, introducing undesirable ambiguity on the complex amplitudes and locations of the sources. The spike deconvolution task amounts to inverting the low-passing effects of the PSF, and to recovering the original sources as precisely as possible.

There is a rich literature on algorithmic investigations of the spike deconvolution problem, ranging from classical root-finding methods such as Prony's method, subspace methods such as MUSIC~\cite{schmidt_multiple_1986,liao2016music}, ESPRIT~\cite{roy_esprit-estimation_1989} and matrix pencil~\cite{moitra2015super}, to more recent optimization methods such as atomic norm minimization  (\emph{a.k.a. total variation minimization})~\cite{de_castro_exact_2012,tang_compressed_2013,candes_towards_2014,chi_harnessing_2019,heckel_super-resolution_2016} and basis pursuit~\cite{malioutov2005sparse,herman_high-resolution_2009}. While classical approaches harness the algebraic properties of complex exponentials by mapping the observations onto a low-dimensional linear subspace to recover the parameters of interest, optimization methods, on the other hand, attempt to recover the parameters via minimizing some carefully-designed loss function. As such, optimization methods tend to be more versatile in adapting to different imaging modalities, as well as amenable to modern advances in large-scale optimization. Inspired by the development of compressive sensing \cite{candes_near-optimal_2006,donoho_compressed_2006}, initial approaches for spike deconvolution relies on a discretization of the spike locations, and then attempts to recover a sparse solution using sparsity-promoting convex relaxations such as the LASSO~\cite{zhu2012faster,tang_sparse_2013}. However, the fundamental issue of basis mismatch~\cite{chi_sensitivity_2009} inherent to the discretization process may significantly hinder the localization performance, and increasing the grid size to reach finer precision levels leads to higher computational cost. Therefore, there has been a surge of interest in developing provably correct convex programs---such as the atomic norm minimization framework mentioned earlier---for spike deconvolution {\em over the continuum} in recent years, with strong performance guarantees developed under sufficient separations between the point sources~\cite{fernandez-granda_super-resolution_2016,ferreira_da_costa_tight_2018,li_approximate_2018,da2020stable}. Nonetheless, the atomic norm framework requires solving a semidefinite program whose complexity scales at least cubically with respect to the signal length; and therefore is computationally expensive and memory inefficient. In addition, although it is in principle possible to examine the so-called dual polynomial to localize the sources~\cite{candes_towards_2014}, it often boils down to an additional post-processing step on the output of the convex program to recover the source parameters, which may hamper the guarantee of the overall procedure.

Motivated by the recent success of nonconvex methods, especially simple first-order methods, in various signal estimation and machine learning tasks~\cite{chi2019nonconvex,chizat2018global}, we are interested in understanding the efficacy of first-order methods in nonconvex spike deconvolution. In fact, first-order methods have already been popular empirically for spike deconvolution, but little is known about their theoretical underpinnings~\cite{huang2017super}. As a first step, this paper focuses on the local geometry and performance guarantees of recovering the parameters of point sources---including both amplitudes and locations---by minimizing a natural nonconvex least-squares loss function measuring the observation residuals.

\subsection{Observation model}

Formally, we formulate the spike deconvolution problem as follows. Consider a vector of $2r$ parameters $\btheta^\star = {[a_1^\star,\ldots,a_{r}^\star,\tau_1^\star,\ldots,\tau_r^\star]}^{\sT}$, where $a_\ell^\star \in {\mathbb{C}}$ and $\tau_\ell^\star \in \mathbb{R}$ correspond to the complex amplitude and location of the $\ell$-th spike, respectively, $\ell =1,\ldots, r$. Denoting by $\cM$ the set of Radon measures over the reals, we assume that the point source signal $\mu^\star \in \cM$ to resolve is of the form
\begin{equation}
	\mu^\star = \mu(\btheta^\star) = \sum_{\ell=1}^r a_\ell^\star \delta_{\tau_\ell^\star},
\end{equation}
where $\delta_{\tau}$ stands for the Dirac function located at $\tau \in \bbR$. Let us further denote the largest and the smallest amplitude of the spikes as
$$a_{\max}^\star =  \max_{1\leq\ell\leq r} \left\vert a_\ell^{\star} \right\vert  = \norm{\ba^\star}_\infty, \qquad a_{\min}^\star = \min_{1\leq\ell\leq r}  \left\vert a_\ell^{\star} \right\vert  . $$
The dynamic range of the measure $\mu^\star$, an important quantity that will be used repetitively later, is thus defined as $ a_{\max}^\star / a_{\min}^\star $.
Denoting by $g \in L_1(\bbR)$ the PSF, the temporal signal $x \in L_1(\bbR)$ resulting from the convolution of the point source signal $\mu^\star$ and the PSF $g$ reads
\begin{align}
	x(\tau) &= \left(g \ast \mu^\star \right)(\tau) = \sum_{\ell = 1}^r a_\ell^\star g(\tau - \tau_\ell^\star),
\end{align}
where $ \ast  $ denotes the convolution product.

A versatile observation model commonly encountered in practice considers the measurements to be taken from a uniform sampling of the Fourier transform of the temporal signal $x$. Denote by $\cF(\cdot)$ the Fourier transform of a measure $\mu$ lying in $\cM$, given by
\begin{equation}\label{eq:Fourier-transform}
	\cF(\mu)(f) = \int_{\bbR} e^{-i 2\pi f\tau} \rd \mu(\tau), \quad \forall \mu \in \cM, \; \forall f \in \bbR.
\end{equation}
Denote by $G = \cF(g)$ and $X = \cF(x)$ the Fourier transform of $g$ and $x$, respectively. We assume that the PSF is band-limited within the bandwidth $B=1$ so that it constrains no frequency greater than $1/2$, \emph{i.e.} $G(f) = 0$ for $\left\vert f \right\vert > \frac{1}{2}$.\footnote{The normalization $B=1$ is made without loss of generality, up to a rescaling of the source locations $\tau_\ell^{\star}$. Assuming that $\tau_\ell^{\star} \in [-\frac{T}{2}, \frac{T}{2}]$, where $T$ is the length of the observation window, the deconvolution problem only depends on the time-bandwidth product $T\cdot B$~\cite{chi_guaranteed_2016}.} For convenience, we assume that an odd number $N = 2n+1$ of measurements are taken in the Fourier domain, uniformly spaced in the bandwidth $[- \frac{1}{2}, \frac{1}{2}]$. The sampled signal $\bx \in \bbC^N$ writes
\begin{align}\label{eq:observationModel}
 \bx &= \bPhi(\mu^\star) \nonumber \\
 & = \diag(\bg) \left[ \cF(\mu) \left(-\frac{n}{N} \right), \dots, \cF(\mu) \left(\frac{n}{N} \right) \right]^{\top},
\end{align}
where $\bPhi : \cM \to \bbC^N$ represents the observation operator and $\bg \in \bbC^N$ is a vector with generic term $g_k = G \left(\frac{k}{N} \right)$ for $k = -n, \dots, n$. Up to a scaling, we assume $\bg$ to have unit Euclidean norm, \emph{i.e.} $\norm{\bg}_2 = 1$.
The goal of spike deconvolution is thus to recover the measure  $\mu^\star$, or equivalently, the parameter $\btheta^\star$, from $\bx$.

\subsection{Our contributions}

In the rest of this paper, we assume that the model order $r$ is known, and consider a natural nonconvex loss function, which aims to minimize the quadratic loss of the parameters $\btheta = {[ a_1,\ldots,a_{r},\tau_1,\ldots,\tau_r]}^{\sT}$  of the Radon measures, given by
\begin{equation}\label{eq:loss_function}
 \min_{\btheta}	\; \cL(\btheta) = \frac{1}{2} \left\lVert \bPhi(\mu(\btheta)) - \bx \right\rVert_2^2.
\end{equation}
Due to the nonlinear form of the parameters, the loss function $\cL(\btheta) $ is clearly nonconvex. As a first step towards nonconvex spike deconvolution, we are interested in understanding the local geometry of the loss function~\eqref{eq:loss_function} and its implications on the computation efficacy of first-order methods. Without loss of generality, we assume
the autocorrelation function $h$ of the PSF $g$ to be a triangular low pass function, \emph{i.e.}
\begin{equation}
	h(\tau) = \int_{-\infty}^{\infty} f(u+\tau) \overline{f(u)} \mathrm{d}u = C \frac{\sin^2\left(\frac{\pi\tau}{2} \right)}{\left( \frac{ \pi \tau}{2} \right)^2},
\end{equation}
where $C>0$ is an constant, and that their respective Fourier transform $H(f)$ and $G(f)$ are linked through the relation $G(f) = \sqrt{H(f)}$ for all $f\in\bbR$. It comes that
\begin{equation}
	g_k = G \left( \frac{k}{N} \right) = \sqrt{H\left( \frac{k}{N} \right)} = \sqrt{ \frac{1}{n+1} \left( 1 - \frac{\left\vert k \right\vert}{n+1} \right) },
\end{equation}
for all $k = -n, \dots n$ after rescaling with the constraint $\norm{\bg}_2 = 1$.
Note that the triangular low-pass function and the Fejér kernel---its discrete counterpart---play an important role in the deconvolution literature and have been extensively proposed as a convolution kernel to evaluate the norm and distance between Radon measures~\cite{candes_towards_2014}. Additionally, our main results can be re-derived with any other bandlimited PSF $g$ by following analogous reasoning, as long as it autocorrelation $h$ is an \emph{absolutely integrable} function.

Concretely, we propose and analyze preconditioned variants of gradient descent (GD), where the search direction is scaled via some carefully designed preconditioning matrices. Our contributions are summarized as follows.
\begin{itemize}
\item We begin with a simple fixed preconditioner design, which adjusts the learning rates of the locations at a different scale from those of the amplitudes, and show it achieves a linear rate of convergence---in terms of {\em entrywise} errors---when initialized close to the ground truth, as long as the separation between the true spikes is sufficiently large. However, the convergence rate slows down significantly when the dynamic range of the source amplitudes is large.
\item To bridge this issue, we introduce an adaptive preconditioner design, which compensates the learning rates of different sources in an iteration-varying manner based on the current estimate. The adaptive design provably leads to an accelerated convergence rate that is independent of the dynamic range, highlighting the benefit of adaptive preconditioning in nonconvex spike deconvolution.
\end{itemize}

Our result is based on understanding the geometric properties of scaled Hessian matrices via a set of novel summation bounds on the absolute sums of sampled Fejér kernels and higher-order derivatives, which might be of independent interest in other contexts.

\subsection{Related work}

The closest work to ours on recovering spike signals from low-pass observations using nonconvex optimization is~\cite{traonmilin2020basins}. The radius of the basin of attraction for gradient descent is characterized whenever the observation operator satisfies the restricted isometry property over the set of well-separated sparse measures. Although the problem setup is versatile, specializing this result in our context of low-pass measurements yields a convergence region whose size scales inversely with the number of sources, which is pessimistic when the number of sources is large. Moreover, the analysis in~\cite{traonmilin2020basins} focuses on the Euclidean error of the parameters, while we focus on the entrywise error, which is more meaningful for gauging the recovery quality of the point sources. Projected gradient methods, which merge pairs of colliding spikes at each iteration, have been proposed in~\cite{traonmilin2020projected,benard2022fast} but without theoretical convergence guarantees.

 Our work can be viewed as falling into a growing line of research on developing provably efficient nonconvex methods---especially first-order methods---for high-dimensional signal estimation, examples including phase retrieval~\cite{candes2015phase,chen2015solving}, low-rank matrix estimation~\cite{sun2016guaranteed,chi2019nonconvex}, blind (sparse) deconvolution~\cite{li2019rapid,chen2021convex,lee2016blind}, dictionary learning~\cite{sun2016complete,Wright2018random}, multi-channel sparse deconvolution~\cite{shi2021manifold,qu2020exact}, and so on. In particular, the preconditioned gradient methods considered in this paper are motivated by~\cite{tong2021accelerating,tong2022scaling,tong2021low}, which demonstrated that preconditioning can efficiently accelerate the convergence of gradient descent in ill-conditioned low-rank estimation.

\subsection{Notation and paper organization}\label{subsec:organization}

Vectors and matrices are denoted by boldface and capital boldface letters, respectively. Vectors $\bx \in \bbC^N$ with odd dimension $N=2n+1$ are indexed between $-n$ and $n$, so that $\bx = [x_{-n},\dots, x_{n}]^{\top}$ for convenience. Transpose and Hermitian transpose of a vector or a matrix $\bA$ are denoted by $\bA^\top$ and $\bA^\herm$, respectively. Furthermore, the adjoint of the operator $\bPhi$ is denoted by $\bPhi^\ast$. We write $\bm{1}_d$ and $\bm{0}_d$ the all-one and null vector (or matrix) in dimension $d$, respectively. With a slight abuse of notation, we denote by $\left\vert \ba \right\vert$, $\left\vert \ba \right\vert^2$, $\ba^{-1}$ the vector with entries equal to the modulus, the squared modulus, and the inverse of the entries of $\ba$, respectively. The element-wise product between two vectors $\ba$ and $\ba^\prime$ is written as $\ba \odot \ba^\prime$. We denote by $\cD_\ell$ the space of $\ell$-times differentiable functions of the real variable. For any function $h\in \cD_\ell$, we write its $\ell$th derivative $h^{(\ell)}$. We denote by $\left\langle \cdot\, , \, \cdot \right\rangle$ and $\left\langle \cdot\, , \, \cdot \right\rangle_\bbR = \Re \left( \left\langle \cdot\, , \, \cdot \right\rangle \right)$ the usual inner product and real  inner product between Radon measure, respectively. Additionally, we let $\delta^{(\ell)} \in \cM$ be the functional which satisfies
\begin{equation}\label{eq:delta_ell_definition}
	g^{(\ell)}(\tau) = \left\langle \delta^{(\ell)}_\tau, g \right\rangle, \quad \forall g\in \cD_\ell, \forall \tau \in \bbR.
\end{equation}

\noindent\textbf{Fejér kernel} We denote by $F_N(\cdot)$ the normalized Fejér kernel of order $N = 2n+1$ defined by
\begin{align}\label{eq:FejerDefinition}
	F_N(t) & = \frac{1}{n+1}\sum_{k=-n}^n \left(1- \frac{\lvert k \rvert}{n+1}\right)e^{i2\pi k t} \nonumber \\
	&= \begin{cases}\frac{\sin^2 \left(\pi(n+1)t\right)}{(n+1)^2 \sin^2(\pi t)}& \text{if } t \notin \bbZ \\
	1 & \text{otherwise}. \end{cases}
\end{align}
The Fejér kernel is a trigonometric polynomial, hence it is infinitely differentiable. We point out that the second derivative of $F_N(\cdot)$ at the origin satisfies
\begin{align}\label{eq:simBound1}
	F_N^{\pprime}(0) & = \frac{1}{n+1} \sum_{k = -n}^n - 4 \pi^2 k^2 \left( 1 - \frac{\left\vert k \right\vert}{n+1} \right) \nonumber \\
	&= -\frac{2}{3} \pi^2 n(n+2) < 0.
\end{align}
Some of its properties, key to this paper, are derived and discussed in Appendix~\ref{sec:Fejer}. The Fejér kernel plays an important role in the sequel as the Gramian $\bPhi^\ast \bPhi: \cM \to \cM$ of the observation operator $\bPhi$ is a convolution product with $F_N$, \emph{i.e.}
\begin{equation}\label{eq:Gramian_expression}
	\bPhi^\ast \bPhi (\mu) = F_N \ast \mu, \quad \forall \mu \in \cM.
\end{equation}

\noindent\textbf{Wrap-around distance} For any set of $r$ points $\btau = \{\tau_1, \dots, \tau_r\} \subset \mathbb{T}$, we denote by $\Delta(\btau)$ is minimal \emph{wrap-around distance}, defined by
\begin{equation}
	\Delta(\btau) \triangleq \min_{\ell \neq \ell^\prime} \inf_{p \in \bbZ} \left\vert \tau_\ell - \tau_{\ell^\prime} + p \right\vert.
\end{equation}

The rest of this paper is organized as follows. Section~\ref{sec:first-order} starts by defining the preconditioned gradient methods and two of its designs with provable local convergence guarantees, using a fixed preconditioner and an adaptive preconditioner in Section~\ref{subsec:fixed_preconditiong} and Section~\ref{subsec:adaptive}, respectively.  Section~\ref{sec:analysis} provides the analysis of the main theorems by controlling the conditioning of the scaled Hessian matrix of the loss function in a neighborhood of the ground truth. Numerical experiments are provided in Section~\ref{sec:numerical_simulations} to corroborate our findings. Finally, a brief conclusion is drawn in Section~\ref{sec:conclusion}.

\section{How does preconditioning help local convergence?}\label{sec:first-order}

\subsection{Preconditioned gradient descent}
Recognizing that the parameters corresponding to the amplitudes and locations may require different treatments, we consider iterates of preconditioned gradient descent (GD) to recover the ground truth parameters, where the preconditioner can possibly be iteration-varying. Given an initialization point $\btheta_0 \in \bbC^{2r}$, the update sequence of preconditioned GD is obtained by successively moving oppositely along the direction of a \emph{linear transform} of the gradient. More specifically, the update rule reads
\begin{equation}\label{eq:scaled_GD}
	\btheta_{k+1} = \btheta_{k} - \bP_k \nabla \cL (\btheta_k),
\end{equation}
where $\btheta_k = {[\ba_k^{\top}, \btau_k^{\top}]}^{\top}$ is the $k$-th iterate, $\bP_k \in \mathbb{C}^{2r \times 2r}$ is a preconditioning matrix (also called preconditioner) that can vary at each iteration; the choice of $\bP_k$ will be detailed momentarily. Here, it is worth noticing that there are no additional learning rates in~\eqref{eq:scaled_GD}, which can be thought of as already absorbed and set within the preconditioner $\bP_k$. By analogy with the celebrated Newton-Raphson method, which selects $\bP_k = {\nabla^2 \cL(\btheta_k)}^{-1}$ (which might however be computationally expensive), the role of the preconditioning matrix $\bP_k$ is to balance the local optimization landscape towards a quadratic function to improve the convergence rate towards a local minimum over the vanilla gradient method. By basic calculation, the gradient $\nabla \cL (\btheta)$ at point $\btheta = {[ a_1,\ldots,a_{r},\tau_1,\ldots,\tau_r]}^{\sT}$ is given by
\begin{subequations}\label{eq:gradient_expression}
	\begin{align}
		\frac{\rd \cL(\btheta)}{\rd a_j}
		&= \left\langle \bPhi(\delta_{\tau_j}), \bPhi(\mu(\btheta) - \mu(\btheta^\star)) \right\rangle\nonumber\\
		&= \left\langle \delta_{\tau_j}, \bPhi^\ast \bPhi(\mu(\btheta) - \mu(\btheta^\star))  \right\rangle \nonumber \\
		&= \left\langle \delta_{\tau_j}, F_N \ast \left( \mu(\btheta) - \mu(\btheta^\star) \right)  \right\rangle \nonumber\\
		&= \sum_{\ell=1}^r a_\ell F_N(\tau_j - \tau_\ell) - \sum_{\ell=1}^{r} a_\ell^\star F_N(\tau_j - \tau_\ell^\star)
	\end{align}
for $ j=1,\ldots, r$,	and similarly,
\begin{align}
		\frac{\rd \cL(\btheta)}{\rd \tau_j}
		&=\left\langle   \bPhi(a_j \delta^{\prime}_{\tau_j}), \bPhi(\mu(\btheta)- \mu(\btheta^\star))  \right\rangle_{\bbR}\nonumber \\
		&=  \left\langle a_j \delta^{\prime}_{\tau_j}, \bPhi^\ast \bPhi(\mu(\btheta) - \mu(\btheta^\star))  \right\rangle_{\bbR} \nonumber \\
		&= \left\langle  a_j \delta^{\prime}_{\tau_j}, F_N \ast \left( \mu(\btheta) - \mu(\btheta^\star) \right)  \right\rangle_{\bbR}\nonumber \\
		&= \Re \Big( \overline{a_j} \Big( \sum_{\ell=1}^r a_\ell F_N^{\prime}(\tau_j - \tau_\ell) - \sum_{\ell=1}^{r} a_\ell^\star F_N^{\prime}(\tau_j - \tau_\ell^\star) \Big) \Big)
	\end{align}
\end{subequations}
for  $j=1,\ldots, r$.

In this paper, we are particularly interested in preconditioning matrices $\bP_k$ that are diagonally structured, so they do not add computation overhead compared with vanilla gradient methods.
In the sequel, we study the basin of attraction and the convergence rate of preconditioned GD for two different preconditioning strategies. The first consists of selecting a time-invariant, diagonal preconditioning matrix $\bP = \bP_k$ whose role is to judiciously renormalize the learning rates between the amplitudes $\ba \in \bbC^r$ and the locations $\btau \in \bbR^r$ which are of different units. The second strategy seeks to dynamically update the preconditioning matrix $\bP_k$ based on the current iterate to better approximate the inverse of the Hessian matrix around the point $\btheta_k$ and accelerate convergence.

\subsection{Invariant preconditioning}\label{subsec:fixed_preconditiong}

In this section, we seek to recover the ground truth parameter $\btheta^\star$ from an instance of the preconditioned GD algorithm~\eqref{eq:scaled_GD} where the sequence of preconditioning matrices is constant, \emph{i.e.} $\bP = \bP_k$ for all $k \in \bbN$. We fix
\begin{equation}\label{eq:P_fix_definition}
	\bP = \diag\left( \begin{bmatrix} \bm{1}_r \\ -{F_N^{\pprime}(0)}^{-1} A^{-2}\bm{1}_r \end{bmatrix} \right),
\end{equation}
where $A > 0$ is an input parameter that controls the ratio between the learning rate applied to the amplitudes $\ba_k$ of the sources and that applied to the locations $\btau_k$ of the sources throughout the iterative process.

\paragraph{Performance metric} To gauge the performance, we define by $\bS \in \bbC^{2r \times 2r}$ the weighting matrix
\begin{equation}\label{eq:S_definition}
	\bS = \diag \left(\begin{bmatrix}
		{\ba^\star}^{-1} \\ \sqrt{-F_N^\pprime(0)} \bm{1}_r,
	\end{bmatrix}\right),
\end{equation}
where $F_N^\pprime(0)$ is given in~\eqref{eq:simBound1}, and study the convergence properties of preconditioned GD in terms of the infinity norm weighted by the matrix $\bS$, \emph{i.e.}
\begin{equation} \label{eq:int_S}
	\norm{\bS \left(\btheta_k - \btheta^\star \right)}_\infty = \max_j \left\{ \frac{ \left\vert a_{k,j} - a_j^\star \right\vert}{\left\vert a_j^\star \right\vert}  , \, \sqrt{-F_N^\pprime(0)} \left\vert \tau_{k,j} - \tau^\star_j \right\vert \right\}.
\end{equation}
Intuitively, the role of this scaling is to analyze a unitless metric that decorrelates the error with the dynamic range of the sources and with the problem dimension, as we have $\sqrt{-F_N^\pprime(0)} = \mathcal{O}(n)$ and that the error on the source locations is expected to be inversely proportional to the number of observation: $\norm{\btau_k - \btau^\star}_\infty = \mathcal{O}\left(n^{-1} \right)$.

The following theorem establishes the linear convergence of preconditioned GD with a fixed preconditioning matrix $\bP$ whenever the input parameter $A$ is properly set, and the initial point $\btheta_0$ is close enough to the ground truth~$\btheta^\star$, as long as the true spikes are sufficiently separated.

\begin{theorem}[Linear convergence with invariant preconditioner]\label{theo:linear_convergence_fixed}
	Suppose that $n \geq 2$ and that the input parameter $A$ satisfies $\norm{\ba^\star}_{\infty}  \leq \frac{3}{2} A$. Moreover, assume that
	\begin{align}\label{eq:eta_definition}
		\eta := 276.21 \frac{A^2  \norm{\ba^\star}_{\infty} }{{(a_{\min}^\star)}^3} \left( \left(n +1 \right) \Delta(\btau^\star) \right)^{-2} < 1,
	\end{align}
	then if the initial point $\btheta_0   = {[\ba_0^{\top}, \btau_0^{\top}]}^{\top}$ satisfies
	\begin{equation} \label{eq:basin_size}
		\norm{\bS \left( \btheta_0 - \btheta^\star \right)}_\infty \leq \frac{1}{2},
	\end{equation}
	the iterates $\{ \btheta_k \}$ of preconditioned GD~\eqref{eq:scaled_GD} with a fixed preconditioner~\eqref{eq:P_fix_definition} converge towards $\btheta^\star$ according to
	\begin{align}\label{eq:linear-convergence-precond-fixed}
		\norm{\bS \left(\btheta_k - \btheta^\star \right)}_\infty \leq \left(1 - \frac{1}{4} \frac{{(a_{\min}^\star)}^2}{A^2} \left(1 - \eta \right)\right)^k \norm{\bS\left(\btheta_0 - \btheta^\star \right)}_\infty
	\end{align}
for all $ k\in \mathbb{N}$.
\end{theorem}

Theorem~\ref{theo:linear_convergence_fixed} indicates that preconditioned GD admits a linear rate of convergence as long as the separation condition $\Delta(\btau^\star)$ is sufficiently large with respect to the dynamic range, {\em i.e.}
\begin{align}
 \left(n +1 \right) \Delta(\btau^\star)  \gtrsim  \left( \frac{\norm{\ba^\star}_\infty}{a_{\min}^\star} \right)^{3/2}.
\end{align}
Additionally, our finding is independent of the number of sources $r$ of the input measure, both in terms of the size of the basin of attraction (cf.~\eqref{eq:basin_size}) and the convergence rate. Faster convergence rate are achieved for smaller values of the parameter $0 < \eta < 1$, when the separation $\Delta(\btau^\star)$ of the true spikes is larger or the dynamic range $\frac{\norm{\ba^\star}_\infty}{a_{\min}^\star}$ of the amplitudes is smaller. However, even for small values of $\eta$, the convergence rate $\rho$ predicted by Theorem~\ref{theo:linear_convergence_fixed} is lower bounded by
\begin{equation}
	\rho \geq 1 - \frac{1}{4}\left( \frac{a_{\min}^\star}{A}\right)^2 \geq 1 - \frac{9}{16} \left( \frac{a_{\min}^\star}{ \|\ba^{\star}\|_{\infty}} \right)^2.
\end{equation}
This suggests that a high dynamic range will lead to a slow convergence rate, independently of the separation $\Delta(\btau^\star)$. Additionally, the convergence guarantees established in Theorem~\ref{theo:linear_convergence_fixed} demand to adjust the input parameter $A$ as a function of $\norm{\ba^\star}_\infty$, which can be impractical in scenarios with no postulate on the norm of the source amplitudes.

\subsection{Adaptive preconditioning}\label{subsec:adaptive}
In order to mitigate the limitations of the fixed preconditioning strategy presented in Section~\ref{subsec:fixed_preconditiong}, we propose to study an instance of preconditioned GD where the preconditioner $\bP_k$ varies at each iteration and is selected as a function of the current iterate $\btheta_k$. In particular, we fix
\begin{equation}\label{eq:P_fix_vary}
	\bP_k = \diag\left( \begin{bmatrix} \bm{1}_r \\ - {F_N^{\pprime}(0)}^{-1} \left\vert \ba_k \right\vert^{-2} \end{bmatrix} \right).
\end{equation}
Similar to Theorem~\ref{theo:linear_convergence_fixed}, the next theorem guarantees a linear convergence rate of the iterates towards the ground truth $\btheta^\star$, provided a good enough initialization point $\btheta_0$, as long as the true spikes are sufficiently separated.
\begin{theorem}[Linear convergence with adaptive preconditioner]\label{theo:linear_convergence}
	Suppose that $n \geq 2$, and assume that
	\begin{equation}\label{eq:gamma-def}
		\gamma := 11.60 \frac{\norm{\ba}_\infty}{a_{\min}^\star} \left( (n+1)\Delta(\btau^\star) \right)^{-2} < \frac{1}{2},
	\end{equation}
	then if the initial point $\btheta_0 = {[\ba_0^{\top}, \btau_0^{\top}]}^{\top}$ satisfies
	\begin{equation}\label{eq:basin_adaptive}
		\norm{\bS \left( \btheta_0 - \btheta^\star \right)}_\infty \leq 1 - \sqrt{\frac{2}{3}},
	\end{equation}
	the iterates $\{ \btheta_k \}$ of preconditioned GD~\eqref{eq:scaled_GD} with an adaptive preconditioner~\eqref{eq:P_fix_vary} converge towards $\btheta^\star$ according to
	\begin{align}\label{eq:linear-convergence-precond}
		\norm{\bS \left(\btheta_k - \btheta^\star \right)}_\infty \leq \left(\frac{1}{2} + \gamma \right)^k \norm{\bS\left(\btheta_0 - \btheta^\star \right)}_\infty
	\end{align}
	for all $k\in \mathbb{N}$.
\end{theorem}

Theorem~\ref{theo:linear_convergence} guarantees that preconditioned GD with an adaptive preconditioner achieves a {\em constant} linear rate of convergence in a similar basin of attraction, provided that  the separation condition $\Delta(\btau^\star)$ is sufficiently large with respect to the dynamic range, {\em i.e.}
\begin{align}
 \left(n +1 \right) \Delta(\btau^\star)  \gtrsim  \left( \frac{\norm{\ba^\star}_\infty}{a_{\min}^\star} \right)^{1/2},
\end{align}
which is much weaker than the requirement for the case using a fixed preconditioner, as indicated in Theorem~\ref{theo:linear_convergence_fixed}. Consequently, this highlights the benefit of adaptive preconditioning in accelerating the convergence in the presence of high dynamic ranges for nonconvex spike deconvolution.

\begin{remark}
We have not attempted to fully optimize the constants in the above theorems. Therefore, their values are set in a quite pessimistic fashion; see Section~\ref{sec:numerical_simulations} for numerical experiments.
\end{remark}

\section{Analysis} \label{sec:analysis}

This section is devoted to proving the two main results of this paper comprised in Theorem~\ref{theo:linear_convergence_fixed} and Theorem~\ref{theo:linear_convergence}. Before entering the core of the proofs, we first provide some warm-up analysis that will be required in the latter proofs.

\subsection{Preliminaries}\label{sec:preliminaries}

\subsubsection{Contraction of entrywise errors}\label{subsec:contraction}
The convergence analysis of the preconditioned gradient method presented in Theorem~\ref{theo:linear_convergence_fixed} and Theorem~\ref{theo:linear_convergence} calls for understanding of the contraction properties of the sequence $\{ \norm{\bS \left(\btheta_k - \btheta^\star \right)}_\infty \}$.
Starting from the update rule~\eqref{eq:scaled_GD}, leveraging $\nabla\cL( \btheta_\star)=\bm{0}$, and applying the fundamental theorem of calculus, we have
\begin{align}\label{eq:descent_equation}
\bS ( \btheta_{k+1} - \btheta_\star)  & = \bS \left(  \btheta_{k} -   \bP_k \nabla  \cL( \btheta_k) - \btheta_\star \right) \nonumber  \\
& = \bS  ( \btheta_{k}  - \btheta_\star)  - \bS   \bP_k \left( \nabla  \cL( \btheta_k) - \nabla\cL( \btheta_\star) \right) \nonumber \\
& = \bS ( \btheta_{k}  - \btheta_\star ) - \bS   \bP_k  \left( \int_0^1 \nabla^2  \cL \left(\btheta^\star + u \left(\btheta_k - \btheta^\star \right) \right) \rd u \right)  \left(\btheta_{k}  - \btheta_\star \right) \nonumber \\
& = \left[\bI   -  \bS \bP_k \left( \int_0^1 \nabla^2  \cL \left(\btheta^\star + u \left(\btheta_k - \btheta^\star \right) \right) \rd u \right)   \bS^{-1}  \right]  \bS    \left(\btheta_{k}  - \btheta_\star \right) .
\end{align}
Let $\mathcal{S}_k = \left\{  \btheta^\star + u   \left( \btheta_k - \btheta^\star \right)   | u\in[0,1] \right\}$ be the line segment that connects $\btheta^\star$ and $\btheta_k$ in $\bbC^{2r}$, and denote by $\rho_k$ the quantity \begin{align}
	\rho_k \triangleq \max_{\btheta \in \mathcal{S}_k}   \norm{\bI -  \bS \bP_k \bH(\btheta)   \bS^{-1}  }_\infty  ,
\end{align}
where $\bH(\btheta) = \nabla^2 \cL(\btheta)$ is the Hessian of the loss function $\cL$ at point $\btheta$. Continuing to bound~\eqref{eq:descent_equation} yields
\begin{align}\label{eq:gradient_descent_theory}
	\norm{ \bS \left( \btheta_{k+1} - \btheta_\star \right) }_\infty & \leq \norm{\bI -  \bS \bP_k \left( \int_0^1 \bH \left(\btheta^\star + u \left(\btheta_k - \btheta^\star \right) \right) \rd u \right) \bS^{-1} }_\infty \norm{  \bS \left(  \btheta_k - \btheta_\star \right)  }_\infty \nonumber \\
	&  =  	\norm{ \int_0^1 \left( \bI -    \bS \bP_k \bH \left(\btheta^\star + u   \left(\btheta_k - \btheta^\star \right) \right) \bS^{-1}   \right)\rd u   }_\infty \norm{ \bS \left( \btheta_k - \btheta_\star \right)  }_\infty \nonumber \\
	&  \leq 	 \left(\int_0^1 \norm{  \bI -    \bS \bP_k \bH \left(\btheta^\star + u   \left(\btheta_k - \btheta^\star \right) \right) \bS^{-1}}_\infty  \rd u   \right) \norm{ \bS \left( \btheta_k - \btheta_\star \right)  }_\infty \nonumber \\
	& \leq  \max_{\btheta \in \mathcal{S}_k} \left\{ \norm{\bI -   \bS \bP_k \bH(\btheta) \bS^{-1}  }_\infty \right\} \norm{  \bS \left( \btheta_k - \btheta_\star \right) }_\infty \nonumber \\
	& = \rho_k \norm{ \bS \left( \btheta_k - \btheta_\star \right) }_\infty.
\end{align}
Hence, the crux of the convergence analysis is to show that $\rho_k<1$ (and control the size of $\rho_k$) uniformly over the segment $\cS_k$ whenever the point $\btheta_k$ lies in an appropriate region centered around the ground truth $\btheta^\star$. Further analysis towards that goal requires an explicit derivation of the Hessian matrix $\bH(\btheta)$, which is done next.

\subsubsection{Hessian decomposition}\label{subsec:hessian_decomposition}

Recall that $\bH(\btheta) = \nabla^2 \cL(\btheta) \in \mathbb{C}^{2r \times 2r}$ is the Hessian matrix of the loss function $\cL$ in~\eqref{eq:loss_function} at the point $\btheta =[\ba^{\top}, \btau^{\top}]^{\top}$. We decompose $\bH(\btheta)$ as
\begin{equation}
	\bH(\btheta) = \begin{bmatrix}
		 \bH_{a,a}(\btheta) & \bH_{a,\tau}(\btheta) \\
		 \bH_{a,\tau}^{\herm}(\btheta) & \bH_{\tau,\tau}(\btheta)
	\end{bmatrix},
\end{equation}
where each block is of size $r \times r$, with generic terms
\begin{align*}
	{[\bH_{a,a}(\btheta)]}_{(i,j)} &= \frac{\rd^2 f(\btheta)}{\rd a_i \rd a_j}, \qquad
	{[\bH_{a,\tau}(\btheta)]}_{(i,j)} = \frac{\rd^2 f(\btheta)}{\rd a_i \rd \tau_j}, \qquad
	{[\bH_{\tau,\tau}(\btheta)]}_{(i,j)} = \frac{\rd^2 f(\btheta)}{\rd \tau_i \rd \tau_j}
\end{align*}
for $i,j = 1\dots,r$. A direct calculation of the Hessian matrix $\bH(\btheta)$ (see, \emph{e.g.}~\cite{traonmilin2020basins}) yields a decomposition of the form
\begin{equation}\label{eq:H_decomposition}
	\bH(\btheta) = \bG(\btheta) + \bE(\btheta),
\end{equation}
where the terms $\bG(\btheta) \in \bbC^{2r \times 2r}$ and $ \bE(\btheta) \in \bbC^{2r \times 2r}$ are described in the sequel.
\paragraph{Structure of $\bG(\btheta)$}
The matrix $\bG(\btheta)$ can be written as
\begin{equation}\label{eq:G_definition}
	\bG(\btheta) = \diag \left(\begin{bmatrix}
	\bm{1}_r \\
	\sqrt{-F_N^{\pprime}(0)}\bm{a}
	\end{bmatrix} \right)^\herm 	\bD(\btau)
	 \diag \left(\begin{bmatrix}
	\bm{1}_r \\
	\sqrt{-F_N^{\pprime}(0)}\bm{a}
	\end{bmatrix} \right),
\end{equation}
with $\bD(\btau) \in \bbC^{2r \times 2r}$ given with a block structure
\begin{equation}\label{eq:D_definition}
	\bD(\btau)= \begin{bmatrix}
		\bD_0(\btau) & \bD_1(\btau) \\
		\bD_1(\btau)^\herm & \bD_2(\btau)
	\end{bmatrix}.
\end{equation}
The entries of the blocks $\bD_0(\btau)$, $\bD_1(\btau)$, $\bD_2(\btau) \in \bbC^{r\times r}$ are composed of
\begin{subequations} \label{eq:D_expressions}
	\begin{align}
		[ \bm{D}_0(\btau)]_{(i,j)} &= F_N(\tau_i - \tau_j), \\
		[ \bm{D}_1(\btau)]_{(i,j)} &= -F_N^{\prime}(\tau_i - \tau_j) / \sqrt{-F_N^{\pprime}(0)}, \\
		[\bm{D}_2(\btau)]_{(i,j)} &= F_N^{\pprime}(\tau_i - \tau_j) / F_N^{\pprime}(0)
	\end{align}
\end{subequations}
for all $i,j = 1,\dots,r$. As shall be seen, the matrix $\bG(\btheta)$ is a relatively  well-conditioned matrix whose spectrum can be controlled as a function of the separation parameter between the spikes $(n+1) \Delta(\btau^\star)$, the dynamic range of the amplitudes $\frac{\norm{\ba}_{\infty}}{a_{\min}^\star}$, and the distance of $\btheta$ to the ground truth parameter $\btheta^\star$.
\paragraph{Structure of $\bE(\btheta)$}
The matrix $\bE(\btheta)$ is given by the block structure decomposition
\begin{equation}\label{eq:E_definition}
\bE(\btheta) =  	\begin{bmatrix}
	\bm{0}_{r\times r} & {\bE_1(\btheta)} \\
	{\bE_1(\btheta)}^\herm & \bE_2(\btheta)
\end{bmatrix},
\end{equation}
where the entries of $\bE_1(\btheta)$, $\bE_2(\btheta)\in \bbC^{r \times r}$ are given as
\begin{subequations}\label{eq:E_entries}
	\begin{align}
		[\bm{E}_1(\btheta)]_{(i,j)} &= \begin{cases} \left\langle \delta^{\prime}_{\tau_j}, F_N \ast  (\mu(\btheta) - \mu(\btheta^\star)) \right\rangle & i=j; \\ 0 & i\neq j;
		\end{cases}\\
		[\bm{E}_2(\btheta)]_{(i,j)} &= \begin{cases} a_j \left\langle \delta^{\pprime}_{\tau_j}, F_N \ast  (\mu(\btheta) - \mu(\btheta^\star))  \right\rangle & i=j; \\ 0 & i\neq j.
		\end{cases}
	\end{align}
\end{subequations}
It is worth noting that $\bE(\btheta_\star) = \bm{0}$, hence $\bE$ can be interpreted as a perturbation term that grows as $\btheta$ deviates from $\btheta_\star$.

With the above preliminaries, we are now ready to prove the main results of this paper. The following two sections present our convergence analyses for preconditioned GD using a fixed preconditioner and an adaptive preconditioner, respectively.

\subsection{Proof of Theorem~\ref{theo:linear_convergence_fixed}}\label{subsec:proof_fixed}

We recall that $\bP_k = \bP = \diag \left( \begin{bmatrix} \bm{1}_r \\ {-F_N^\pprime(0)}^{-1} A^{-2}\bm{1}_r  \end{bmatrix} \right)$ for all $k \in \bbN$ when the preconditioner is fixed.
The contraction analysis~\eqref{eq:gradient_descent_theory} presented in Section~\ref{subsec:contraction} suggests that the contraction rate in Theorem~\ref{theo:linear_convergence_fixed} is controlled by the quantity $\norm{\bS \bP \bH(\btheta) \bS^{-1} - \bI}_\infty$ in neighborhood around the ground truth $\btheta^\star$. The next theorem, whose proof is deferred to Appendix~\ref{sec:proof_uniformBounds}, provides a uniform bound on this quantity on a neighborhood of the ground truth.

\begin{theorem}[Uniform bound of the Hessian] \label{theo:uniformBounds_fixed} Suppose that $n \geq 2$, $(n+1) \Delta(\btau^\star) \geq 16.5$ and $\norm{\btau - \btau^\star}_\infty \leq \frac{1}{4} \Delta(\btau^\star)$. Let $\btheta_k = [\ba_k^{\top}, \btau_k^{\top}]^{\top}$ and assume that $A \geq \max \left\{ \frac{3}{2} \norm{\ba^\star}_\infty, \; \norm{\ba_k}_\infty \right\}$  then the exist two positive constants
	\begin{subequations}\label{eq:K_values}
		\begin{align}
			K_{\Delta} & = 2.13, \\
			K_\theta & = 44.42,
		\end{align}
	\end{subequations}
	such that for all $\btheta = [\ba^{\top}, \btau^{\top}]^{\top} \in \cS_k$ with $\norm{\bm{S} \left(\btheta_k - \btheta^\star \right)}_\infty < 1$,
	we have that
	\begin{multline}
		\norm{\bS \bP \bH(\btheta) \bS^{-1} - \bI}_\infty \leq 1 - \left(\frac{a_{\min}^\star}{A}\right)^2 \left(1 - \norm{ \bS(\btheta_k - \btheta^\star)}_\infty \right)^2  \\
			+\left( 4 K_\Delta  + K_\theta \norm{\bS \left( \btheta_k - \btheta^\star \right)}_\infty \right)  \frac{\| \ba^\star\|_{\infty}}{a_{\min}^\star} \left( (n+1) \Delta(\btau^\star)\right)^{-2} \left( 1 + \norm{\bS \left(\btheta_k - \btheta^\star \right)}_\infty \right)^2 .
	\end{multline}
\end{theorem}
Theorem~\ref{theo:uniformBounds_fixed} provides a bound on $\norm{\bS \bP \bH(\btheta) \bS^{-1} - \bI}_\infty $ depending on the quantity $\frac{\| \ba^\star\|_{\infty}}{a_{\min}^\star} \left( (n+1) \Delta(\btau^\star) \right)^{-2}$, which can be made small enough under the hypothesis of Theorem~\ref{theo:linear_convergence_fixed}.
We proceed with the rest of the proof by induction.

For the base case, it is trivial that~\eqref{eq:linear-convergence-precond-fixed} holds for $k=0$. We start the induction by assuming that
\begin{align} \label{eq:induction_hypothesis}
	\norm{\bS \left(\btheta_k - \btheta^\star \right)}_\infty & \leq \left( 1 - \frac{1}{4} \left(\frac{a_{\min}^\star}{A}\right)^2(1-\eta) \right)^k \norm{\bS \left(\btheta_0 - \btheta^\star \right)}_\infty
\end{align}
holds for some $k \in \bbN$. We begin by verifying the assumptions of Theorem~\ref{theo:uniformBounds_fixed}.
\begin{itemize}
\item First, as the dynamic range $\frac{\norm{\ba^\star}_\infty}{a^\star_{\min}} \geq 1$, it is easy to see that the hypothesis~\eqref{eq:eta_definition} immediately implies that $(n+1) \Delta(\btau^\star) \geq 16.6$.
\item By the definition~\eqref{eq:int_S} and the induction hypothesis~\eqref{eq:induction_hypothesis}, it follows that
\begin{align}
	\norm{\btau_k - \btau^\star}_\infty &\leq \frac{1}{\sqrt{-F_N^\pprime(0)}} \norm{\bS \left(\btheta_k - \btheta^\star \right)}_\infty \nonumber \\
	&\leq \frac{1}{\sqrt{-F_N^\pprime(0)}} \norm{\bS \left(\btheta_0 - \btheta^\star \right)}_\infty \nonumber \\
	&\leq \frac{\sqrt{3}}{2 \pi} \frac{1}{n+1}  \leq \frac{1}{n+1} \leq \frac{\Delta(\btau^\star)}{4}.
\end{align}
\item Furthermore, we have that
\begin{align}
	\norm{\ba_k}_\infty &\leq \norm{\ba^\star}_\infty + \norm{\ba_k - \ba^\star}_\infty \nonumber \\
	&\leq \norm{\ba^\star}_\infty \left( 1 + \norm{\bS \left( \btheta_k - \btheta^\star \right)}_\infty \right) \nonumber \\
	&< \norm{\ba^\star}_\infty \left( 1 + \norm{\bS \left( \btheta_0 - \btheta^\star \right)}_\infty \right) \nonumber \\
	&\leq \frac{3}{2} \norm{\ba^\star}_\infty \leq A.
\end{align}
\end{itemize}
Hence, the assumptions of Theorem~\ref{theo:uniformBounds_fixed} hold, which yields
\begin{align}\label{eq:H_bound_dev}
	\rho_k & = \max_{\btheta \in \cS_k} \norm{\bS \bP \bH(\btheta) \bS^{-1} - \bI}_\infty \nonumber \\
	&\leq 1 - \frac{{(a_{\min}^\star)}^2}{4 A^2} + \frac{9}{4}\left( 4 K_\Delta  + \frac{1}{2} K_\theta \right)  \frac{{\| \ba^\star\|}_{\infty} }{a_{\min}^\star} \left( (n+1) \Delta(\btau^\star)\right)^{-2} \nonumber \\
	&\leq 1 - \frac{{(a_{\min}^\star)}^2}{4 A^2}\left(1 - 9\left(4 K_\Delta + \frac{1}{2} K_\theta \right) \frac{A^2 {\| \ba^\star\|}_{\infty} }{{a_{\min}^\star}^3} \left( (n+1) \Delta(\btau^\star)\right)^{-2}  \right) \nonumber \\
	&\leq 1 - \frac{{(a_{\min}^\star)}^2}{4 A^2}\left(1 - 276.21 \frac{A^2 {\| \ba^\star\|}_{\infty} }{{a_{\min}^\star}^3} \left( (n+1) \Delta(\btau^\star)\right)^{-2}  \right) \nonumber \\
	&\leq 1 - \frac{{(a_{\min}^\star)}^2}{4 A^2} \left(1 - \eta \right),
\end{align}
where we substituted the definition~\eqref{eq:eta_definition} of $\eta$ in the last line. It results from the iterative analysis~\eqref{eq:gradient_descent_theory} that the next update $\btheta_{k+1}$ obeys
\begin{align}
	\norm{\bS \left(\btheta_{k+1} - \btheta^\star \right)}_\infty & \leq \rho_k \norm{\bS \left(\btheta_k - \btheta^\star \right)}_\infty \nonumber\\
	&\leq  \left(1 - \frac{{(a_{\min}^\star)}^2}{4 A^2} \left(1 - \eta \right) \right) \norm{\bS \left(\btheta_k - \btheta^\star \right)}_\infty \nonumber\\
	&\leq \left(1 - \frac{{(a_{\min}^\star)}^2}{4 A^2} \left(1 - \eta \right) \right)^{k+1} \norm{\bS \left(\btheta_{0} - \btheta^\star \right)}_\infty,
\end{align}
which concludes the proof of Theorem~\ref{theo:linear_convergence_fixed}.

\subsection{Proof of Theorem~\ref{theo:linear_convergence}}\label{subsec:proof_adaptive}

We proceed with the proof of Theorem~\ref{theo:linear_convergence} analogously to the proof of Theorem~\ref{theo:linear_convergence_fixed} presented in Section~\ref{subsec:proof_fixed}. First, we establish the following intermediate theorem that controls the conditioning of the scaled Hessian matrix $\bS\bP_k(\btheta)\bH\bS^{-1}$ uniformly over the segment $\cS_k$ as a function of the weighted infinity-norm distance $\norm{\bS \left( \btheta_k - \btheta^\star \right)}_\infty$. The proof of Theorem~\ref{theo:uniformBounds} is deferred to Appendix~\ref{subsec:proof_uniformBounds_adaptive}.

\begin{theorem}[Uniform bound of the Hessian] \label{theo:uniformBounds} Suppose that $n \geq 2$, $(n+1) \Delta(\btau^\star) \geq 4.7$ and $\norm{\btau - \btau^\star}_\infty \leq \frac{1}{4} \Delta(\btau^\star)$ then the exists two positive constants
	\begin{subequations}\label{eq:K_values_adap}
		\begin{align}
			K_{\Delta} & \leq 2.32 ,\\
			K_\theta & \leq 75.80,
		\end{align}
	\end{subequations}
	such that for all $\btheta = [\ba^{\top}, \btau^{\top}]^{\top} \in \cS_k$ satisfying $\norm{\bm{S} \left(\btheta_k - \btheta^\star \right)}_\infty < 1$
	we have that
	\begin{multline} 
		\norm{\bS \bP_k \bH(\btheta) \bS^{-1} - \bI}_\infty \leq \frac{1}{\left(1 - \norm{ \bS(\btheta_k - \btheta^\star)}_\infty \right)^2} - 1 \\
		+ \left(4 K_\Delta + K_\theta \norm{\bS \left(\btheta_k - \btheta^\star \right)}_\infty  \right) \frac{ \| \ba^\star \|_{\infty}}{a_{\min}^\star} \frac{\left( (n+1) \Delta(\btau^\star)\right)^{-2}}{\left( 1 - \norm{\bS \left(\btheta_k - \btheta^\star \right)}_\infty \right)^2}.
	\end{multline}
\end{theorem}
The rest of the  proof follows similarly by induction.
For the base case, it is trivial that the initial point $\btheta_0$ verifies~\eqref{eq:linear-convergence-precond}. We now assume that $\btheta_k$ satisfies
\begin{align} \label{eq:induction_hypothesis_adaptive}
	\norm{\bS \left(\btheta_k - \btheta^\star \right)}_\infty & \leq \left(\frac{1}{2} + \gamma \right)^k \norm{\bS \left(\btheta_0 - \btheta^\star \right)}_\infty
\end{align}
for some $k \in \bbN$. Let's verify the assumptions of Theorem~\ref{theo:uniformBounds}.
\begin{itemize}
\item First, as the dynamic range $\frac{\norm{\ba^\star}_\infty}{a^\star_{\min}} \geq 1$, the assumption~\eqref{eq:gamma-def} easily implies that $(n+1)\Delta(\btau^\star) \geq 4.7$.
\item By the definition~\eqref{eq:int_S} and the induction hypothesis~\eqref{eq:induction_hypothesis_adaptive}, it follows that
\begin{align}
	\norm{\btau_k - \btau^\star}_\infty &\leq \frac{1}{\sqrt{-F_N^\pprime(0)}} \norm{\bS \left(\btheta_k - \btheta^\star \right)}_\infty \nonumber \\
	&\leq \frac{1}{\sqrt{-F_N^\pprime(0)}} \norm{\bS \left(\btheta_0 - \btheta^\star \right)}_\infty \nonumber \\
	&\leq \frac{\sqrt{3}}{\pi} \left(1 - \sqrt{\frac{2}{3}} \right) \frac{1}{n+1} \nonumber \\
	& \leq \frac{1}{n+1} \leq \frac{\Delta(\btau^\star)}{4}.
\end{align}
\end{itemize}
Hence, the assumptions of Theorem~\ref{theo:uniformBounds} hold. Noticing that the function $ f( u) :=   \frac{1}{\left(1-u\right)^2}$ is increasing over $[0,1)$, we have that $\frac{1}{\left(1 - \norm{\bS \left(\btheta_k - \btheta^\star \right)^2}_\infty \right)} \leq \frac{1}{2}$.  Together with Theorem~\ref{theo:uniformBounds}, this yields the bound
\begin{align}\label{eq:H_bound_dev_adap}
	\rho_k  &= \max_{\btheta \in \cS_k} \norm{\bS \bP_k \bH(\btheta) \bS^{-1} - \bI}_\infty \nonumber \\
	&\leq \frac{1}{2} +   \left(4 K_\Delta + K_\theta \left( 1 - \sqrt{\frac{2}{3}} \right) \right) \frac{ \| \ba^{\star}\|_{\infty}}{2 a_{\min}^\star} \left( (n+1) \Delta(\btau)\right)^{-2} \nonumber\\
	&\leq \frac{1}{2} + 11.60 \frac{\| \ba^{\star}\|_{\infty}}{a_{\min}^\star} \left( (n+1) \Delta(\btau^\star)\right)^{-2} \nonumber\\
	&\leq \frac{1}{2} + \gamma < 1,
\end{align}
where we substituted the definition~\eqref{eq:gamma-def} of $\gamma$ in the third inequality. It results from the iterative analysis~\eqref{eq:gradient_descent_theory} that the next update $\btheta_{k+1}$ satisfies
\begin{align}
	\norm{\bS \left(\btheta_{k+1} - \btheta^\star \right)}_\infty \leq& \rho_k  \norm{\bS \left(\btheta_k - \btheta^\star \right)}_\infty \nonumber\\
	\leq&   \left(\frac{1}{2} + \gamma \right)  \norm{\bS \left(\btheta_k - \btheta^\star \right)}_\infty \nonumber\\
	\leq&   \left(\frac{1}{2} + \gamma \right)^{k+1} \norm{\bS \left(\btheta_{0} - \btheta^\star \right)}_\infty,
\end{align}
which concludes the proof of Theorem~\ref{theo:linear_convergence}.

\section{Numerical experiments}\label{sec:numerical_simulations}

This section provides a numerical validation of Theorem~\ref{theo:linear_convergence_fixed} and Theorem~\ref{theo:linear_convergence}. In the following experiments, the signal length is set to $N = 65$ (\emph{i.e.} $n=32$). The ground truth signal is composed of $r = 6$ sources placed in the interval $[-\frac{1}{2}, \frac{1}{2})$ while ensuring that $(n+1) \Delta(\btau^\star) \geq 2$, which is a more optimistic separation condition than what the theorems' statements suggest. Additionally, the dynamic range is denoted by $\kappa = \frac{\norm{\ba^\star}_\infty}{a_{\min}^\star}$. The complex amplitudes $\ba^\star \in \bbC^{r}$ are selected independently and uniformly at random in a complex annulus with bounds $1 \leq |a_\ell^\star| \leq \kappa$. The input parameter of the invariant preconditioning scheme is set at $A = \frac{3}{2} \norm{\ba^\star}_\infty$.

\begin{figure}[ht]
	\centering
	\begin{tabular}{cc}
	\includegraphics[width=0.47\textwidth]{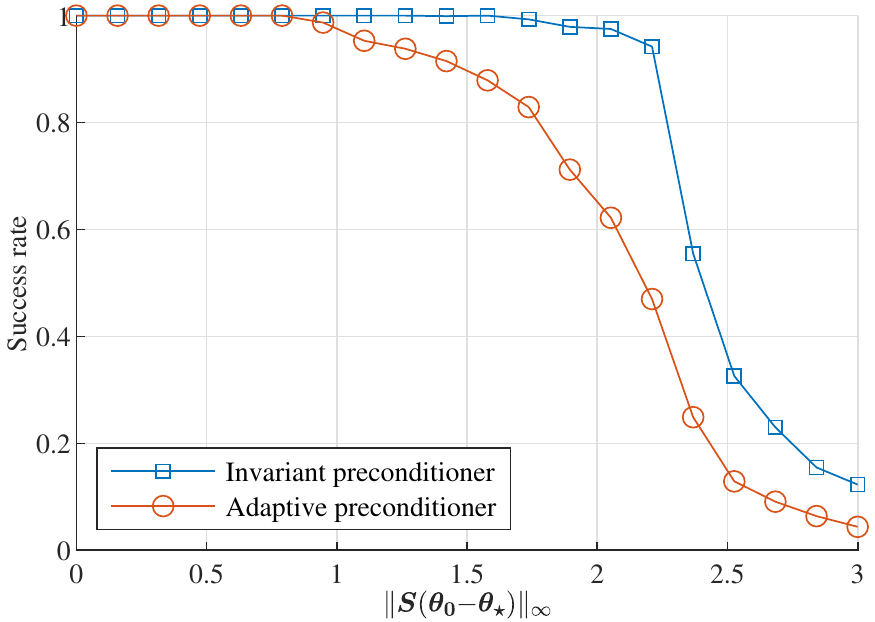}&
	\includegraphics[width=0.47\textwidth]{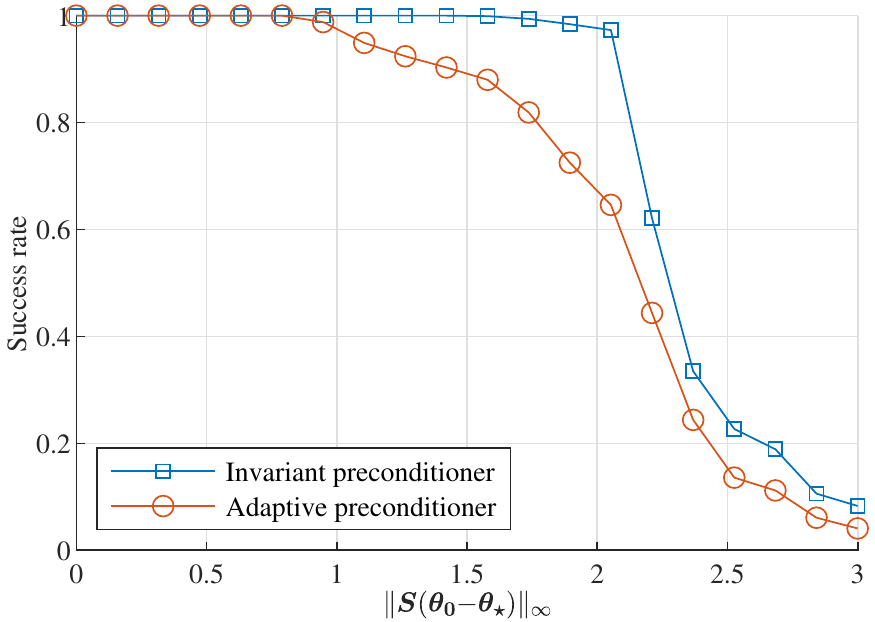} \\
	(a) $\kappa = 1$ & (b) $\kappa =  6$
	\end{tabular}
	\caption{Success rate of the invariant and adaptive preconditioning schemes on reconstructing the ground truth $\btheta_\star$ as a function of the initialization distance $\norm{\bS \left(\btheta_0 - \btheta_\star \right)}_\infty$ for different dynamic ranges $\kappa$. The results are averaged over 1000 randomized trials.}
	\label{fig:size_basin}
\end{figure}

\paragraph{Size of the basin of attraction}Of critical importance in the analysis of Theorem~\ref{theo:linear_convergence_fixed} and Theorem~\ref{theo:linear_convergence} is the distance between the initial parameter $\btheta_0$ and the ground truth $\btheta_\star$.
We start by comparing the success rates of both preconditioning schemes on reconstructing the ground truth as a function of the initialization distance $\norm{\bS \left( \btheta_0 - \btheta_\star \right)}_\infty$. In each experiment, the starting point $\btheta_0$ is drawn uniformly over the set of points equidistant to $\btheta_\star$. An experiment is labeled as a success if $\norm{\bS \left( \btheta_{200} - \btheta_{\star} \right)}_\infty \leq 10^{-2}$ after 200 iterations. Figure~\ref{fig:size_basin} suggests that, for both schemes, the size of the basin of attraction is independent of the dynamic range $\kappa$, and is around the order of magnitude $\norm{\bS \left( \btheta_0 - \btheta_\star \right)}_\infty \simeq 1$. This suggests that the numerical constants ($1/2$ and $1-\sqrt{2/3}\sim 0.184$, respectively) set forth in Theorem~\ref{theo:linear_convergence_fixed} and Theorem~\ref{theo:linear_convergence}, respectively, are pessimistic and nonconvex spike deconvolution performs in a much more benign manner than predicted by our theory, indicating room for further refinements.

\paragraph{Linear convergence using a spectral initialization}
In practice, several ad hoc initialization methods could be envisaged to produce an initial point $\btheta_0$ that falls in the basin of attraction of the preconditioned gradient descent methods. Herein, we proceed by uniformly discretizing the spectral domain over $N$ elements. Given the knowledge of the ground truth model order $r$, the initial locations $\btau_0$ are selected as the $r$ elements of the discrete grid whose weighted Fourier transform best describes the observation $\bx$. Mathematically,  consider the following optimization problem
\begin{align}\label{eq:initialization_opt}
	\bu_0 &= \argmin_{\bu} \; \frac{1}{2}\norm{\bx - \diag(\bg) \bF_N \bu}_2^2 \;\; \text{s.t.}\; \norm{\bu}_0 \leq r,
\end{align}
where $\bF_N \in \bbC^{N \times N}$ is a discrete Fourier transform matrix, and the $\ell_0$-norm denotes the cardinality of the support. Writing $\cI_0 = \mbox{supp}\left(\bu_0 \right) \subset [-n, \dots, n]$ the support of the solution $\bu_0$ of~\eqref{eq:initialization_opt}, the parameter $\btheta_0$ is constructed in a second stage by selecting $\btau_0 = {[\frac{k_1}{N}, \dots, \frac{k_{r}}{N}]}^\top$ where $k_\ell \in \cI_0$, $\ell = 1,\dots,r$ and $\ba_0 = \bu_{|\cI_0}$ as the restriction of $\bu$ to the elements in $\cI_0$.
The program~\eqref{eq:initialization_opt} is itself a non-convex sparse reconstruction problem, which we approximate the solution using the orthogonal matching pursuit algorithm~\cite{tropp2007signal}. The proposed initialization procedure offers several benefits over more classical methods: It is highly scalable, robust to high dynamic range, and does not involve any polynomial root finding subroutine.

\begin{figure}[ht]
	\centering
	\begin{tabular}{cc}
	\includegraphics[width=0.47\textwidth]{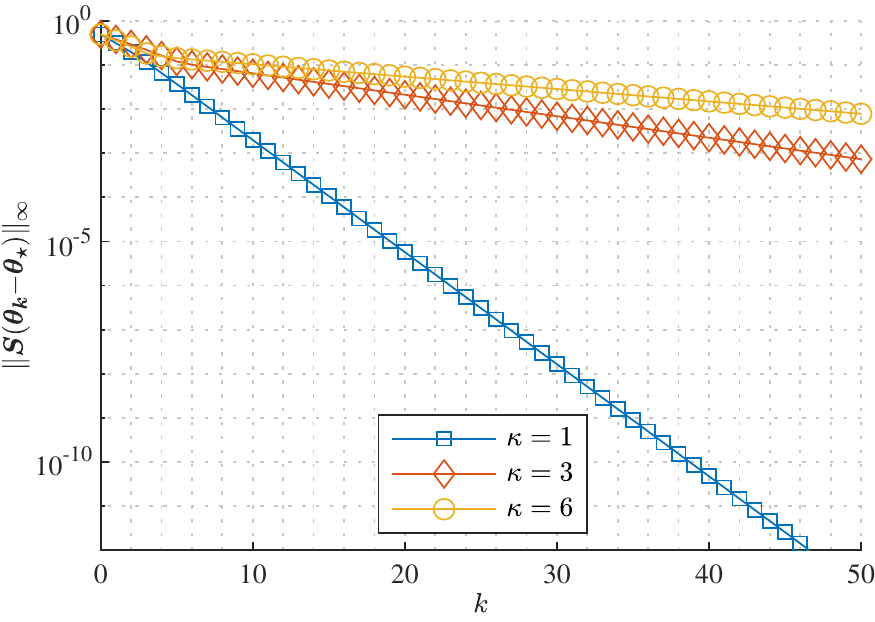}&
	\includegraphics[width=0.47\textwidth]{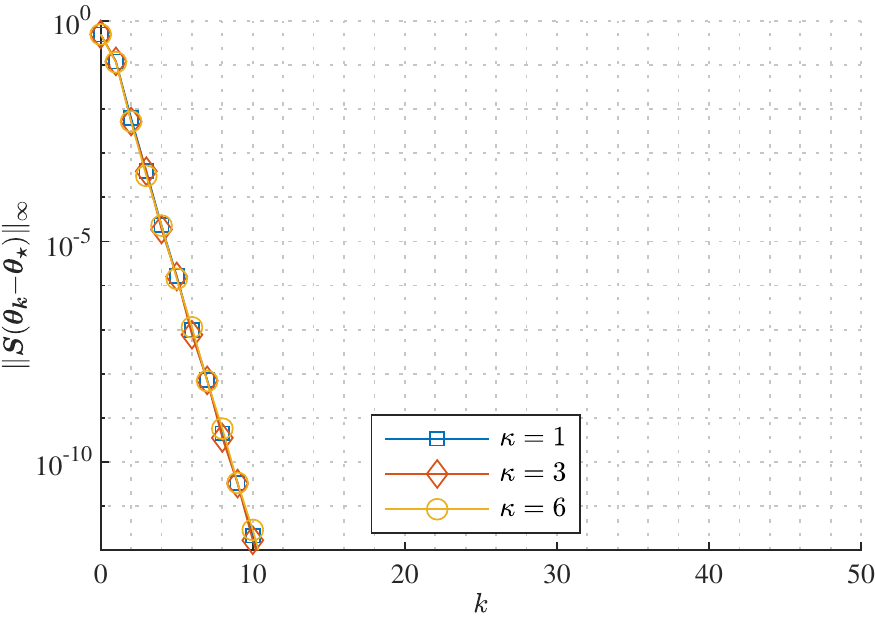} \\
	(a) Invariant preconditioning & (b) Adaptive preconditioning
	\end{tabular}
	\caption{Convergence rates of the iterate sequence of preconditioned GD towards the ground truth as the dynamic range $\kappa$ varies for: (a) the invariant preconditioning scheme; (b) the adaptive preconditioning scheme.}
	\label{fig:convergence_dynamic_range}
\end{figure}

Figure~\ref{fig:convergence_dynamic_range} pictures the convergence rate of preconditioned GD under the invariant and adaptive preconditioning schemes, respectively. For both schemes, $\btheta_0$ is selected according to the previously described initialization procedure. It can be seen that, although both preconditioning schemes ensure a linear converge of the iterate sequence, the convergence rate with a fixed preconditioner degrades as the dynamic range of the sources increases. In contrast, the one with an adaptive preconditioner remains unchanged. Additionally, the adaptive preconditioning scheme benefits from faster convergence rates for a given dynamic range. These experimental results corroborate the theoretical findings presented in Section~\ref{sec:first-order}.

\begin{figure}[t]
	\centering
	\includegraphics[width=0.85\columnwidth]{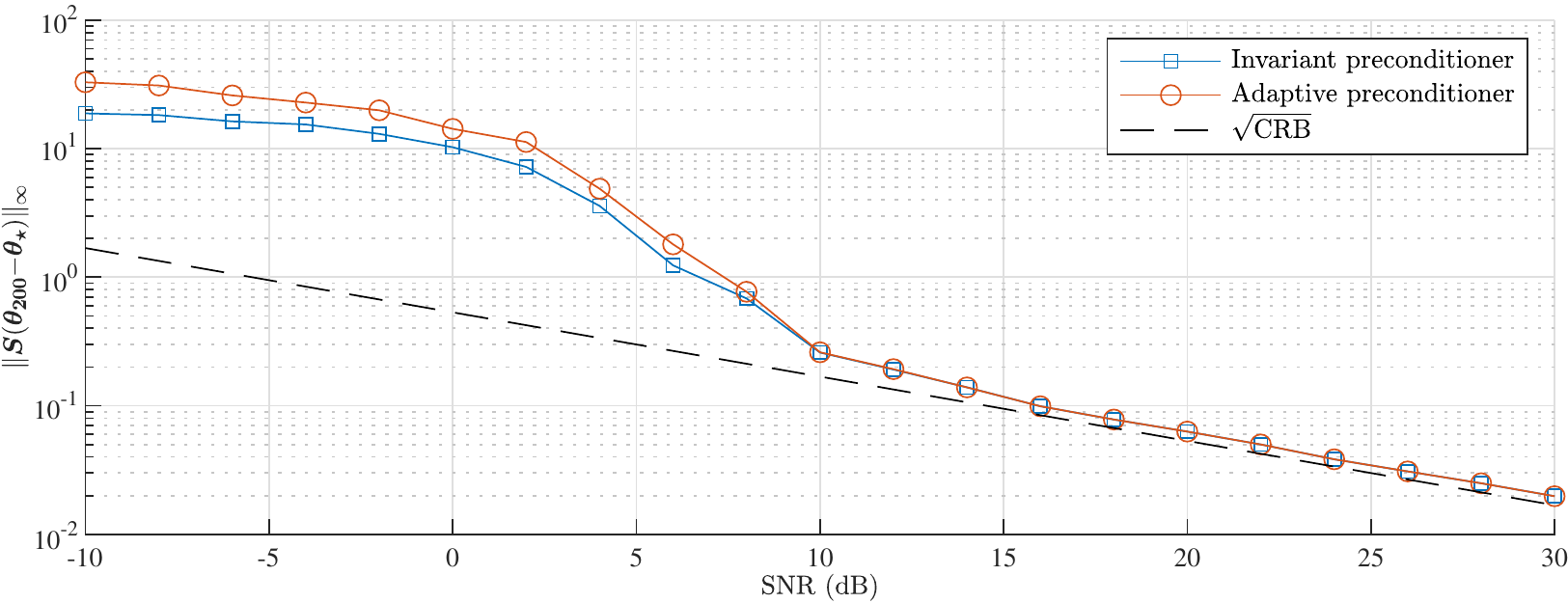}
	\caption{Statistical error $\norm{\bS \left( \btheta_{200} - \btheta_{\star} \right)}_\infty$ of both preconditioned GD schemes as a function of the SNR. The dynamic range is set to $\kappa =3$, and the results are averaged over 1000 randomized trials.}
	\label{fig:perf_vs_SNR}
\end{figure}

\paragraph{Noisy recovery} We next examine the performance of preconditioned GD in the presence of noise. We assume observations of the form $\bx = \bm{\Phi}(\mu^\star) + \bw$, where $\bw$ is white Gaussian noise, and estimate $\mu^\star$ by minimizing~\eqref{eq:loss_function} starting from an initial point $\btheta_0$ obtained by the spectral initialization procedure described above.
Figure~\ref{fig:perf_vs_SNR} draws the statistical error $\norm{\bS \left( \btheta_{200} - \btheta_{\star} \right)}_\infty$ of both preconditioning schemes after 200 iterations --- when convergence is reached --- as a function of the signal-to-noise ratio (SNR), defined as $\mathsf{SNR} = \norm{\bm{\Phi}(\mu^\star)}_2^2/\norm{\bm{w}}_2^2$. The results are benchmarked against the Cramér-Rao bound (CRB) \cite{scharf1993geometry}. Both statistical errors remain close to the CRB under a sufficiently large SNR, providing an empirical validation of the robustness of the proposed algorithms.

\section{Conclusion}\label{sec:conclusion}

This work proposed and analyzed preconditioned gradient methods for nonconvex spike deconvolution using both fixed and adaptive preconditioners, and demonstrated that for ground truth with sufficiently separated spikes, the proposed methods achieve a linear rate of convergence that is independent of the number of spikes, as long as a close enough initialization is provided near the ground truth. In particular, by designing the preconditioner to compensate adaptively for the amplitude profile of the spikes, it is possible to accelerate the convergence rate to be dimension-free and independent of the dynamic range, while the convergence using a fixed preconditioner slows down when the dynamic range is large. Our work thus highlights the importance of preconditioning in accelerating convergence in nonconvex spike deconvolution.

As a first step towards understanding the efficacy of first-order methods for spike deconvolution, this works opens up several interesting directions for further investigation.
\begin{itemize}
\item {\em Initialization schemes.} One immediate direction is to analyze initialization schemes that produce initial estimates that fall into the basin of attraction, which we suspect the procedure described in Section~\ref{sec:numerical_simulations} is a good candidate.
\item {\em Model order.} For simplicity, it is assumed that the model order $r$ is known perfectly, which might not hold in practice. It is of great interest to develop modified algorithms when the model order is overspecified, which has recently been examined comprehensively in \cite{xu2023power} for low-rank estimation from small random initializations.
\item {\em General observations.} Another direction is to extend the analysis to more general observation operators, possibly including random sampling, missing data, as well as corruptions. This may necessarily require a reformulation of the loss function, such as a nonsmooth and nonconvex formulation using the least absolute deviation~\cite{tong2021low} to improve robustness.
\item {\em Separation condition.} Last but not least, it is of great importance to study to what extent it is possible to relax the success condition in terms of the separation condition, possibly with additional positive constraints of the source amplitudes.
\end{itemize}

\appendix

\section{Summation bounds of the Fejér kernel} \label{sec:Fejer}

The purpose of this section is to present Lemma~\ref{lem:BoundsFejerKernel}, which delivers fundamental bounds on the absolute sum of the Fejér kernel and its derivatives at sampled points of interest. Although specific to the Fejér kernel, Lemma~\ref{lem:BoundsFejerKernel} could be adapted to any other absolutely integrable point spread function without a significant change in the proof structure.

\begin{lemma}[Uniform bounds on the Fejér kernel]\label{lem:BoundsFejerKernel}
	Suppose that $n \geq 2$. Let $\btau = \{\tau_1, \dots, \tau_r\} \subset \bbT$, and let $\alpha > 0$ be such that $(n+1) \Delta(\btau) \geq \alpha$. Let $\bu = \{u_{i,j}\}_{i \neq j} \subset \bbR$ be a set of $\frac{r(r-1)}{2}$ real numbers that are absolutely bounded by $\beta$ such that
	\begin{equation}
		(n+1) \max_{i\neq j} \{\left\vert u_{i,j} \right\vert\} \triangleq \beta < \frac{\alpha}{2}.
	\end{equation}
	Then the inequalities
	\begin{align}\label{eq:boundsFejerKernel}
		\max_{i} \sum_{j\neq i} \left\vert F_N^{(\ell)}(\tau_j - \tau_i + u_{i,j}) \right\vert & \leq  C_\ell {(n+1)}^{\ell} \left( (n+1)\Delta(\btau)\right)^{-2}
	\end{align}
	hold for $\ell = 0,1,2,3$, where the constants $C_\ell$ only depend on $\alpha$ and $\beta$ and are given by
	\begin{subequations}\label{eq:constant_values}
		\begin{align}
			C_0 & =  \frac{4}{\pi^2}\left(\alpha - 2\beta \right)^{-1} \alpha, \\
			C_1 &= \left(\frac{4}{\pi}\left(\alpha - 2\beta \right)^{-1}  + \frac{8}{\pi^2} \left(\alpha - 2\beta \right)^{-2} \right) \alpha, \\
			C_2 &= \left(\frac{80}{9}\left(\alpha - 2\beta \right)^{-1} + \frac{16}{\pi} \left(\alpha - 2\beta \right)^{-2}   + \frac{64}{3 \pi^2} \left(\alpha - 2\beta \right)^{-3} \right)\alpha, \\
			C_3 &=  \left( \frac{16}{3}\left(\alpha - 2\beta \right)^{-1} \pi + \frac{1488}{27} \left(\alpha - 2\beta \right)^{-2}   + \frac{192}{\pi} \left(\alpha - 2\beta \right)^{-3} + \frac{192}{\pi^2} \left(\alpha - 2\beta \right)^{-4}  \right)\alpha.
		\end{align}
	\end{subequations}

\end{lemma}

\newcommand{\sinpt}{{\sin \left( (n+1) \pi t \right)}}
\newcommand{\cospt}{{\cos \left( (n+1) \pi t \right)}}
\newcommand{\costwopt}{{\cos^2 \left( (n+1) \pi t \right)}}
\newcommand{\sintwopt}{{\sin^2 \left( (n+1) \pi t \right)}}


\begin{proof} First, the function $F_N$ is a trigonometric polynomial, hence infinitely differentiable. Let us begin by examining the general expression of $F_N(t)$ and its derivatives up to the third order. Assuming $t \notin \bbZ$, by basic calculation, we have that
	\begin{subequations}\label{eq:FejerDerivatives}
		\begin{align}
			F_N(t) & = \frac{\sin^2 \left(\pi(n+1)t\right)}{(n+1)^2 \sin^2(\pi t)} , \\
			F_N^\prime(t) & =  \frac{2\pi}{n+1} \cos((n+1) \pi t) \sin((n+1) \pi t) \csc^2(\pi t) \nonumber \\
			& \qquad - \frac{2 \pi}{(n+1)^2} \sin((n+1)\pi t) \cot(\pi t) \csc^2(\pi t) , \\
			F_N^\pprime(t) & =  2 \pi^2 \left( \cos^2((n+1) \pi t) - \sin^2\left((n+1) \pi t\right) \right) \csc^2 (\pi t) \nonumber \\
			{}& \qquad -\frac{8 \pi^2}{n+1} \cos((n+1) \pi t) \sin((n+1) \pi t) \cot(\pi t) \csc^2(\pi t) \nonumber\\
			{}& \qquad + \frac{2\pi^2}{(n+1)^2}
			\sin^2(\pi(n+1)t)\left(2\cot^2(\pi t) + 1 \right) \csc^2(\pi t) \\
			F_N^\ppprime(t) & =  -(n+1) 8 \pi^3  \cospt \sinpt \csc^2(\pi t) \nonumber \\
			{}& \qquad - 12 \pi^3 \left( \costwopt - \sintwopt \right) \cot(\pi t) \csc^2(\pi t) \nonumber \\
			{}& \qquad + \frac{12 \pi^3}{n+1} \cospt \sinpt \left(3 \cot^2(\pi t)
			+1 \right)\csc^2(\pi t) \nonumber \\
			{}& \qquad - \frac{8 \pi^3}{\left(n + 1 \right)^2} \sintwopt \left(3 \cot^3(\pi t) + 2 \cot(\pi t)  \right)\csc^2(\pi t).
		\end{align}
	\end{subequations}
Using the four trigonometric bounds $\left\vert \sin(a) \right\vert \leq 1$, $\left\vert \cos(a) \right\vert \leq 1$, $\left\vert \cos(a) \sin(a) \right\vert \leq \frac{1}{2}$, and $\left\vert \cos^2(a) - \sin^2(a) \right\vert \leq 1$ for $a \in \bbR$, and the triangle inequality,~\eqref{eq:FejerDerivatives} can be further absolutely bounded as
	\begin{subequations}\label{eq:absolute_bound_FejerDerivatives}
		\begin{align}
			\left\vert F_N(t) \right\vert & \leq  \frac{1}{(n+1)^2} \csc^2(\pi t) , \\
			\left\vert F_N^\prime(t) \right\vert & \leq  \frac{\pi}{n+1} \csc^2(\pi t) + \frac{2 \pi}{(n+1)^2} \left\vert \cot(\pi t) \right\vert \csc^2(\pi t), \\
			\left\vert F_N^\pprime(t) \right\vert & \leq 2\pi^2 \csc^2(\pi t) + \frac{4\pi^2}{n+1} \left\vert \cot(\pi t) \right\vert \csc^2(\pi t) + \frac{2 \pi^2}{(n+1)^2} \left(2\cot^2(\pi t) + 1 \right)\csc^2 (\pi t)\\
			\left\vert F_N^\ppprime(t) \right\vert & \leq (n+1)4 \pi^3 \csc^2(\pi t) + 12 \pi^3 \left\vert \cot(t) \right\vert \csc^2(t) + \frac{12 \pi^3}{n+1}\left(3 \cot^2(\pi t)
			+1 \right)\csc^2(\pi t) \nonumber \\
			{}& \qquad + \frac{8 \pi^3}{\left(n+1 \right)^2} \left(3 \left\vert \cot^3(\pi t) \right\vert +  2 \left\vert\cot(\pi t) \right\vert  \right) \csc^2(\pi t).
 		\end{align}
	\end{subequations}
To continue, observe that a basic building block in the bound~\eqref{eq:absolute_bound_FejerDerivatives} takes the following function form, which is denoted by
\begin{align}\label{eq:hell}
		h_\ell(t) & =  \cot^\ell(\pi t) \csc^2(\pi t), \qquad \ell = 0, 1, 2.
\end{align}
The functions $h_\ell$, $\ell=0,1,2$ are even, $1$-periodic, and  continuous, non-negative, decreasing and convex over the interval $(0,\frac{1}{2}]$. Define the three sums $S_\ell(n,\btau,\bu,i)$, $\ell=0,1,2$ by
	\begin{equation}\label{eq:defSl}
		S_\ell(n,\btau,\bu,i) = \sum_{j \neq i} h_\ell \left( \tau_j - \tau_i + u_{i,j} \right).
	\end{equation}
It is straightforward to see that the terms of interest can be bounded in terms of $S_\ell(n,\btau,\bu,i)$ in view of~\eqref{eq:absolute_bound_FejerDerivatives}. We shall claim the following bound of $S_\ell(n,\btau,\bu,i)$ holds, which will be proven at the end of proof:
	\begin{align}\label{eq:boundsOnSl}
		S_\ell(n,\btau,\bu,i) \leq \left(\frac{2}{\pi \Delta(\btau)} \right)^{\ell + 2} \frac{1}{\left(\ell + 1 \right) \left(1- 2 \alpha^{-1}\beta  \right)^{\ell + 1}}, \qquad \ell=0,1,2.
	\end{align}
Interestingly, the above bound on $S_\ell\left(n,\btau,\bu,i \right)$ does not depend on $n$ or $i$.
We are now ready to establish the bound~\eqref{eq:boundsFejerKernel} of interest. First, we have
	\begin{align}
		\max_{i} \sum_{j\neq i} \left\vert F_N(\tau_j - \tau_i + u_{i,j}) \right\vert &  \leq  \sum_{j\neq i} \frac{1}{(n+1)^2} h_0(\tau_j - \tau_i + u_{i,j}) \nonumber\\
		&=  \frac{1}{(n+1)^2} S_0(n,\btau,\bu,i) \nonumber \\
		&\leq  \frac{4}{\pi^2} \frac{{\Delta(\btau)}^{-2}}{ (n+1)^2 \left(1-2  \alpha^{-1}\beta \right)}\nonumber \\
		&= \frac{4}{\pi^2}\left(\alpha - 2\beta \right)^{-1} \alpha \left( (n+1)\Delta(\btau)\right)^{-2}.
	\end{align}
Similarly, for the case of the first derivative, we have
\begin{align}
\max_{i} \sum_{j\neq i} \left\vert F_N^\prime(\tau_j - \tau_i + u_{i,j}) \right\vert & \leq \sum_{j\neq i}\left(   \frac{\pi}{n+1} h_0(\tau_j - \tau_i + u_{i,j})+  \frac{2 \pi}{{(n+1)}^2} h_1(\tau_j - \tau_i + u_{i,j}) \right)  \nonumber\\
		&= \frac{\pi}{n+1} S_0(n,\btau,\bu,i) + \frac{2\pi}{(n+1)^2} S_1(n,\btau,\bu,i)\nonumber \\
		& \leq \frac{4 {\Delta(\btau)}^{-2}}{\pi (n+1) \left(1-2\alpha^{-1}\beta \right)} + \frac{8{\Delta(\btau)}^{-3}}{\pi^2 {(n+1)}^2 \left( 1-2\alpha^{-1}\beta \right)^2} \nonumber \\
		& \leq \left(\frac{4}{\pi}\left(\alpha - 2\beta \right)^{-1}  + \frac{8}{\pi^2} \left(\alpha - 2\beta \right)^{-2} \right) \alpha{(n+1)} \left( (n+1)\Delta(\btau)\right)^{-2},
\end{align}
where the last line uses $(n+1) \Delta(\btau) \geq \alpha$. Moving onto the second derivative, it follows
\begin{align}
 	&\max_{i} \sum_{j\neq i} \left\vert F_N^\pprime(\tau_j - \tau_i + u_{i,j}) \right\vert \nonumber\\
	& \leq \sum_{j\neq i} \left(  2\pi^2 h_0(\tau_j - \tau_i + u_{i,j})+  \frac{4 \pi^2}{n+1} h_1(\tau_j - \tau_i + u_{i,j})  + \frac{2\pi^2}{{(n+1)}^2} \left( 2 h_2 (\tau_j - \tau_i + u_{i,j}) + h_0 (\tau_j - \tau_i + u_{i,j})  \right) \right) \nonumber\\
			& =  2\pi^2 S_0(n,\btau,\bu,i) + \frac{4\pi^2}{n+1} S_1(n,\btau,\bu,i)+ \frac{2\pi^2}{{(n+1)}^2} \left( 2 S_2(n,\btau,\bu,i) + S_0(n,\btau,\bu,i) \right)\nonumber \\
			&\leq 2\pi^2 \left(1 + \frac{1}{\left(n+1\right)^2} \right) S_0(n,\btau,\bu,i) + \frac{4\pi^2}{n+1} S_1(n,\btau,\bu,i)+ \frac{4\pi^2}{{(n+1)}^2}   S_2(n,\btau,\bu,i) \nonumber \\
			& \leq  \frac{20\pi^2}{9}  S_0(n,\btau,\bu,i) + \frac{4\pi^2}{n+1} S_1(n,\btau,\bu,i)+ \frac{4\pi^2}{{(n+1)}^2}   S_2(n,\btau,\bu,i) \nonumber \\
			&\leq{} \frac{80 {\Delta(\btau)}^{-2}}{9 \left(1-2\alpha^{-1}\beta\right)} + \frac{16 {\Delta(\btau)}^{-3}}{\pi (n+1) \left( 1-2\alpha^{-1}\beta \right)^2} + \frac{64 {\Delta(\btau)}^{-4}}{3 \pi^2 {(n+1)}^2 \left(1-2\alpha^{-1}\beta \right)^{3}}  \nonumber \\
			&\leq{} \left(\frac{80}{9}\left(\alpha - 2\beta \right)^{-1} + \frac{16}{\pi} \left(\alpha - 2\beta \right)^{-2} + \frac{64}{3 \pi^2} \left(\alpha - 2\beta \right)^{-3}\right)\alpha{(n+1)}^{2} \left( (n+1)\Delta(\btau)\right)^{-2}.
	\end{align}
	Finally, for the third derivative, it holds
	\begin{align}
	& \max_{i} \sum_{j\neq i} \left\vert F_N^\ppprime(\tau_j - \tau_i + u_{i,j}) \right\vert  \nonumber \\
		&\leq \sum_{j\neq i} \Big(   (n+1)4\pi^3 h_0(\tau_j - \tau_i + u_{i,j})  +  12 \pi^3 h_1(\tau_j - \tau_i + u_{i,j}) \nonumber \\
		&  \qquad+ \frac{12 \pi^3}{{n+1}} \left( 3 h_2 (\tau_j - \tau_i + u_{i,j}) + h_0 (\tau_j - \tau_i + u_{i,j})  \right) + \frac{8 \pi^3}{{(n+1)}^2} \left(3 h_3(\tau_j - \tau_i + u_{i,j}) + 2 h_1(\tau_j - \tau_i + u_{i,j}) \right) \Big) \nonumber \\
		 & =  (n+1) 4\pi^3 S_0(n,\btau,\bu,i) + 12\pi^3  S_1(n,\btau,\bu,i) + \frac{12 \pi^3}{n+1} \left(3  S_2(n,\btau,\bu,i) +  S_0(n,\btau,\bu,i) \right)  \nonumber\\
		 &  \qquad + \frac{8 \pi^3}{{(n+1)}^2} \left(3  S_3(n,\btau,\bu,i) +  2 S_1(n,\btau,\bu,i) \right) \nonumber \\
		&\leq  (n+1) 4\pi^3 \left( 1 + \frac{3}{\left(n+1\right)^2} \right) S_0(n,\btau,\bu,i) + 12\pi^3 \left(1 + \frac{4}{3 \left(n+1\right)^2}\right) S_1(n,\btau,\bu,i) \nonumber\\
		 & \qquad + \frac{36 \pi^3}{n+1}   S_2(n,\btau,\bu,i)  + \frac{24 \pi^3}{{(n+1)}^2} S_3(n,\btau,\bu,i)\nonumber\\
		& \leq (n+1) \frac{64}{3} \pi \frac{{\Delta(\btau)}^{-2}}{1-2\alpha^{-1} \beta} + \frac{2488}{27} \frac{{\Delta(\btau)}^{-3}
		}{\left( 1-2\alpha^{-1}\beta \right)^2}   + \frac{192}{\pi (n+1)} \frac{{\Delta(\btau)}^{-4}}{\left( 1-2 \alpha^{-1} \beta \right)^3} + \frac{192}{\pi^2 {(n+1)}^2}\frac{{\Delta(\btau)}^{-5}}{\left( 1-2 \alpha^{-1} \beta \right)^4} \nonumber \\
		&\leq  \left( \frac{16}{3}\left(\alpha - 2\beta \right)^{-1} \pi + \frac{1488}{27} \left(\alpha - 2\beta \right)^{-2} + \frac{192}{\pi} \left(\alpha - 2\beta \right)^{-3} + \frac{192}{\pi^2} \left(\alpha - 2\beta \right)^{-4} \right)   \alpha{(n+1)}^{3} \left( (n+1)\Delta(\btau)\right)^{-2}.
	\end{align}
	The proof is thus completed if we can prove~\eqref{eq:boundsOnSl}, which is the focus of the rest of the proof.

\paragraph{Proof of~\eqref{eq:boundsOnSl}} We fix the index $i$ and take the convention $u_{i,i} = 0$. As the functions ${\{h_\ell\}}$ are 1-periodic, the quantity $S_\ell(n,\btau,\beta,i)$ is invariant by integer translations of $\{ \tau_j \}$'s. Therefore, one can make the assumption, up to a modulo considerations and a reordering of the indices that the sequence $\{\tau_j - \tau_i + u_{i,j} \}_j$ is within the range $[-\frac{1}{2}, \frac{1}{2})$ and in an ascending order, so that
\begin{equation*}
	-\frac{1}{2} \leq \tau_1 - \tau_i + u_{i,1} < \tau_2 - \tau_i + u_{i,2} < \dots < \tau_r - \tau_i + u_{i,r} < \frac{1}{2}.
\end{equation*}
Denote by $r_+$ and $r_-$ the number of positive and negative elements in the set $\left\{ \tau_j - \tau_i + u_{i,j} \right\}_{j}$, respectively. As $\tau_j - \tau_i + u_{i,j} = 0$ if and only if $i=j$, we have that $ r_+ + r_{-} = r - 1$.
Using the separation condition, and as $h_\ell$ is decreasing over $(0,\frac{1}{2}]$ and even, we have that
		\begin{align*}
			0 \leq h_\ell \left(\tau_{i+j} - \tau_{i} + u_{i,j} \right) \leq h_\ell \left( j\Delta(\btau) - \frac{\beta}{n+1} \right),&\quad j = 1,\dots, r_+;\\
		 0 \leq h_\ell \left( \tau_{i-j} - \tau_{i} + u_{i,j} \right) \leq{} h_\ell \left( -j \Delta(\btau) + \frac{\beta}{n+1} \right) = h_\ell \left( j \Delta(\btau) - \frac{\beta}{n+1} \right),&\quad j = 1,\dots, r_{-}.
	 \end{align*}
We can subsequently bound the sum~\eqref{eq:defSl} as
	\begin{align}\label{eq:SlBound1}
		S_\ell(n,\btau,\bu,i) ={}& \sum_{j = 1}^{r_+} h_\ell \left( \tau_{i+j} - \tau_i + u_{i,(i+j)} \right) + \sum_{j = 1}^{r_-} h_\ell \left( \tau_{i-j} - \tau_i + u_{i,(i-j)} \right) \nonumber \\
		\leq{}& \sum_{j = 1}^{r_+} h_\ell \left( j \Delta(\btau) - \frac{\beta}{n+1} \right) + \sum_{j = 1}^{r_-} h_\ell \left( j \Delta(\btau) - \frac{\beta}{n+1} \right) \nonumber \\
		\leq{}& 2 \sum_{j=1}^{\left\lceil \frac{r-1}{2} \right\rceil} h_\ell \left( j \Delta(\btau) - \frac{\beta}{n+1} \right).
	\end{align}
	Identifying and associating the right-hand side of~\eqref{eq:SlBound1} to a Riemann sum with a mid-point rule and recalling that $h_\ell$ is decreasing and convex over $(0,\frac{1}{2}]$ leads to the majorant
	\begin{align}\label{eq:SlRiemannSum}
		S_\ell(n,\btau,\bu,i) \leq{}& 2 \int_{\frac{\Delta(\btau)}{2} - \frac{\beta}{n+1}}^{\left(\left\lceil \frac{r-1}{2} \right\rceil + \frac{1}{2} \right) \Delta(\btau) - \frac{\beta}{n+1}} h_\ell(u) \rd u \nonumber \\
		\leq{}& 2 \int_{\frac{\Delta(\btau)}{2} - \frac{\beta}{n+1}}^{\frac{1}{2}}  h_\ell(u) \rd u \nonumber \\
		={}& 2 \left( H_\ell\left(\frac{1}{2} \right) - H_\ell \left( \frac{\Delta(\btau)}{2} - \frac{\beta}{n+1} \right) \right),
	\end{align}
	where we used the non-negativity of $h_\ell$ and the inequality $\left(\left\lceil \frac{r-1}{2} \right\rceil + \frac{1}{2} \right) \Delta(\btau) - \frac{\beta}{n+1} \leq \frac{1}{2}$ in the second inequality. Here, $\{H_\ell\}$'s are primitives of the functions $\{ h_\ell\}$ over the interval $(0,\frac{1}{2}]$. Moreover, we have
		\begin{align}\label{eq:hPrimitives}
			H_\ell(t) ={}& -\frac{1}{\pi \left( \ell + 1\right)} \cot^{\ell + 1}(\pi t) + c
		\end{align}
	for $\ell = 0, 1, 2$, where $c$ is an arbitrary constant. This yields, with the inequality $0 \leq \cot(\pi u) \leq \frac{1}{\pi u}$ for all $0<u<\frac{1}{2}$, a further simplification of~\eqref{eq:SlRiemannSum}:
	\begin{align}
		\Delta(\btau) S_\ell(n,\btau,\bu,i) \leq{}& \frac{2}{\pi(\ell + 1)} \cot^{\ell + 1} \left( \pi \left( \frac{\Delta(\btau)}{2} - \frac{\beta}{n+1} \right) \right) \nonumber \\
		\leq{}& \left( \frac{2}{\pi} \right)^{\ell + 2} \Delta(\btau)^{-\ell - 1} \frac{1}{\left(1-\frac{2 \beta {\Delta(\btau)}^{-1}}{n+1} \right)^{\ell + 1}} \nonumber \\
		\leq{}& \left( \frac{2}{\pi} \right)^{\ell + 2} \Delta(\btau)^{-\ell - 1} \frac{1}{\left(1-2 \alpha^{-1} \beta \right)^{\ell + 1}},
	\end{align}
which immediately leads to the claimed bound~\eqref{eq:boundsOnSl}.
\end{proof}

\section{Proof of the uniform Hessian bounds} \label{sec:proof_uniformBounds}

This section is dedicated to establish Theorem~\ref{theo:uniformBounds_fixed} and Theorem~\ref{theo:uniformBounds}.

\subsection{Technical lemmas}\label{subsec:uniformBounds_preliminary_lemmas}

Lemma~\ref{lem:boundsOnD} and Lemma~\ref{lem:boundOnEDev} provide bounds on core quantities that are involved in the decomposition of the Hessian matrix $\bH(\btheta)$ characterized in Section~\ref{subsec:hessian_decomposition}. Their proofs are presented in Appendix~\ref{subsec:proof_boundOnD} and Appendix~\ref{subsec:proof_boundOnE}, respectively.

\begin{lemma}\label{lem:boundsOnD}
	Suppose that $n \geq 2$ and let $\btau \subset \bbT$ be such that $(n+1)\Delta(\btau) \geq \alpha$ for some $\alpha > 0$, then there exists a constant $K_{\Delta}$ with
	\begin{equation}
		K_{\Delta} := \max \left\{ C_0 + \frac{3\sqrt{3}}{4\pi} C_1, \frac{3\sqrt{3}}{4\pi} C_1 + \frac{27}{16 \pi^2} C_2 \right\}
	\end{equation}
	where the constants $C_0,C_1,C_2 > 0$ are defined in in~\eqref{eq:constant_values} with parameters $\alpha$ and $\beta = 0$ such that
	\begin{equation}
		\left\Vert \bD(\btau) - \bI \right\Vert_\infty \leq K_{\Delta} ((n+1)\Delta(\btau))^{-2}.
	\end{equation}
\end{lemma}

\begin{lemma}\label{lem:boundOnEDev}
	Suppose that $n \geq 2$ and let $\btau_\star = [ \tau_1^\star ,\dots,\tau_r^\star ]^\top$ and $\btau = [ \tau_1,\dots, \tau_r ]^\top$ be two vectors of points around the torus. Assume that $(n+1)\Delta(\btau^\star) \geq \alpha$. As long as $\left(n+1\right) \left\Vert \btau - \btau^\star \right\Vert_\infty \leq \beta < \frac{\alpha}{2}$, we have
	\begin{subequations}
		\begin{align}
			\left\vert \left\langle \Phi(\delta^{\prime}_{\tau_j}), \Phi \left(\mu(\btheta)) - \mu(\btheta_\star) \right) \right\rangle \right\vert &  \leq \left(C_1 \left\Vert \ba - \ba^\star \right\Vert_\infty + C_2  \left\Vert \ba^\star \right\Vert_\infty (n+1)\left\Vert \btau - \btau^\star \right\Vert_\infty \right) (n+1) \left( (n+1) \Delta(\btau)\right)^{-2}, \\
			\left\vert \left\langle \Phi(\delta^{\pprime}_{\tau_j}), \Phi \left(\mu(\btheta)) - \mu(\btheta_\star) \right) \right\rangle \right\vert & \leq \left(C_2 \left\Vert \ba - \ba^\star \right\Vert_\infty + C_3  \left\Vert \ba^\star \right\Vert_\infty (n+1)\left\Vert \btau - \btau^\star \right\Vert_\infty \right) {(n+1)}^2  \left( (n+1) \Delta(\btau)\right)^{-2}
		\end{align}
	\end{subequations}
	for all $j = 1, \dots, r$, where the constants $C_1, C_2, C_3 > 0$ are defined in~\eqref{eq:constant_values} with parameters $(\alpha,\beta)$.
\end{lemma}

\subsection{Proof of Theorem \ref{theo:uniformBounds_fixed}}\label{subsec:proof_uniformBounds_fixed}

Recalling the expression of the Hessian in~\eqref{eq:H_decomposition}, it follows that
\begin{equation}
	\bS \bP \bH(\btheta) \bS^{-1}    - \bI  = \bS \bP \bG(\btheta) \bS^{-1} - \bI   +  \bS \bP \bE(\btheta) \bS^{-1}.
\end{equation}
We proceed to bound $\left\|\bS \bP \bG(\btheta) \bS^{-1} - \bI\right\|_{\infty}$ and $\left\| \bS \bP \bE(\btheta) \bS^{-1} \right\|_\infty$ separately, and then combine them via the triangle inequality.

\paragraph{Step 1: bound $\left\| \bS \bP \bG(\btheta) \bS^{-1}    - \bI  \right\|_\infty$}
From~\eqref{eq:S_definition} and~\eqref{eq:G_definition}, we have that
\begin{align}
	  \bS \bP \bG(\btheta) \bS^{-1}    - \bI
	& = \bS \bP \diag \left(\begin{bmatrix}
		\bm{1}_r \\
		\sqrt{-F_N^\pprime(0)} \ba
		\end{bmatrix} \right)^\sH 	\bD(\btau)
		\diag \left(\begin{bmatrix}
			\bm{1}_r \\
			\sqrt{-F_N^\pprime(0)} \ba
			\end{bmatrix} \right)  \bS^{-1} - \bI \nonumber\\
	& = \diag \left(\begin{bmatrix}
			{\ba^\star}^{-1} \\
			{ A^{-2}}  \ba
			\end{bmatrix} \right)^\sH 	\bD(\btau)
			\diag \left(\begin{bmatrix}
			\ba^\star \\
			\ba
			\end{bmatrix} \right) - \bI \nonumber \\
			& = \diag \left(\begin{bmatrix}
				{\ba^\star}^{-1} \\
			 	{A^{-2}}  \ba
				\end{bmatrix} \right)^\sH 	\left( \bD(\btau) - \bI \right)
				\diag \left(\begin{bmatrix}
				\ba^\star \\
				\ba
				\end{bmatrix} \right) + \diag \left(\begin{bmatrix}
					\bm{1}_r \\
					{A^{-2}}  \left\vert \ba \right\vert^2
					\end{bmatrix} \right) - \bI.
\end{align}
This immediately yields from the triangle inequality that
\begin{align}\label{eq:bG_bound_fix_before}
	\MoveEqLeft \norm{ \bS \bP \bG(\btheta) \bS^{-1}    - \bI}_\infty & \nonumber \\
	& \leq \norm{\diag \left(\begin{bmatrix}
		{\ba^\star}^{-1} \\
		{ A^{-2}} \odot \ba
		\end{bmatrix} \right)^\sH 	\left( \bD(\btau) - \bI \right)
		\diag \left(\begin{bmatrix}
		\ba^\star \\
		\ba
		\end{bmatrix} \right)}_\infty + \norm{\diag \left(\begin{bmatrix}
			\bm{1}_r \\
			{A^{-2}} \odot \left\vert \ba \right\vert^2
			\end{bmatrix} \right) - \bI}_\infty \nonumber \\
	& \leq \norm{\diag \left(\begin{bmatrix}
		{\ba^\star}^{-1} \\
		{A^{-2}} \odot \ba
		\end{bmatrix} \right)^\sH}_\infty \norm{	  \bD(\btau) - \bI   }_\infty
		\norm{ \diag \left(\begin{bmatrix}
		\ba^\star \\
		\ba
		\end{bmatrix} \right)}_\infty + \norm{\diag \left(\begin{bmatrix}
			\bm{1}_r \\
			{A^{-2}} \odot \left\vert \ba \right\vert^2
			\end{bmatrix} \right) - \bI}_\infty \nonumber \\
	&\leq \max_j \left\{\frac{1}{\left\vert a_j^\star \right\vert}, \frac{ \left\vert a_{j} \right\vert}{A^2} \right\} \max_j \left\{ \left\vert a_j^\star \right\vert , \left\vert a_j  \right\vert \right\} \norm{  \bD(\btau) - \bI  }_\infty + \max_j \left\{ \left\vert 1 - \frac{\left\vert a_j \right\vert^2}{A^2} \right\vert \right\}.
	\end{align}
	The three maxima in~\eqref{eq:bG_bound_fix_before} can be controlled using the basic relation
	\begin{align}\label{eq:dynamic_err}
		1 - \norm{\bS (\btheta_k - \btheta^\star)}_\infty \leq \frac{\left\vert a_j^\star \right\vert - \left\vert a_{k,j} - a_j^\star \right\vert}{\left\vert a_j^\star \right\vert} \leq \frac{\left\vert a_j \right\vert}{\left\vert a_{j}^\star \right\vert} &\leq \frac{\left\vert a_j^\star \right\vert + \left\vert a_{k,j} - a_j^\star \right\vert}{\left\vert a_j^\star \right\vert} \leq 1 + \norm{\bS (\btheta_k - \btheta^\star)}_\infty,
	\end{align}
	for $j = 1,\dots,r$ and by exploiting the assumption $\|\ba^{\star}\|_{\infty}\leq A$ as follows
	\begin{subequations}\label{eq:A-S-bounds-1}
		\begin{align}
			\max_j \left\{\frac{1}{\left\vert a_j^\star \right\vert}, \frac{ \left\vert a_{j} \right\vert}{A^2} \right\} &\leq \max_j \left\{\frac{1}{\left\vert a_j^\star \right\vert}, \frac{ \left\vert a_{j}^\star \right\vert}{A^2} \left(1 + \norm{\bS (\btheta_k - \btheta^\star)}_\infty  \right) \right\} \leq \frac{1}{a_{\min}^\star} \left(1 + \norm{\bS (\btheta_k - \btheta^\star)}_\infty  \right) ,\\
			\max_j \left\{ \left\vert a_j^\star \right\vert , \left\vert a_j  \right\vert \right\} &\leq \max_j \left\{ \left\vert a_j^\star \right\vert , \left\vert a_j^\star \right\vert \left( 1 + \norm{\bS (\btheta_k - \btheta^\star)}_\infty \right)  \right\} \leq \norm{\ba^\star}_\infty \left( 1 + \norm{\bS (\btheta_k - \btheta^\star)}_\infty \right) ,\\
			\max_j \left\{ \left\vert 1 - \frac{\left\vert a_j \right\vert^2}{A^2} \right\vert \right\} &\leq 1 - \min_j \left\{  \frac{\left\vert a_j \right\vert^2}{A^2}  \right\} \leq 1 - \frac{\left(a^\star_{\min}\right)^2}{A^2}\left( 1 - \norm{\bS (\btheta_k - \btheta^\star)}_\infty\right).
		\end{align}
	\end{subequations}
The bounds~\eqref{eq:A-S-bounds-1} and Lemma~\ref{lem:boundsOnD} imply with~\eqref{eq:bG_bound_fix_before} that
	\begin{align}\label{eq:bG_bound_fix}
		\MoveEqLeft \norm{ \bS \bP \bG(\btheta) \bS^{-1}    - \bI}_\infty& \nonumber \\
		&\leq \frac{ \norm{\ba^\star}_{\infty} }{a_{\min}^\star} \left(1 + \norm{\bS \left( \btheta_k - \btheta^\star \right)}_\infty \right)^2 \norm{   \bD(\btau) - \bI   }_\infty + 1 - \frac{{ (a_{\min}^\star)^2 \left( 1 - \norm{\bS \left( \btheta_k - \btheta^\star \right)}_\infty \right)}^2}{A^2} \nonumber\\
	&\leq \frac{ \norm{\ba^\star}_{\infty} }{a_{\min}^\star}\left(1 + \norm{\bS \left( \btheta_k - \btheta^\star \right)}_\infty \right)^2  K_\Delta \left((n+1) \Delta(\btau)\right)^{-2} + 1 - \frac{{ (a_{\min}^\star)^2 \left( 1- \norm{\bS \left( \btheta_k - \btheta^\star \right)}_\infty \right)}^2}{A^2},
\end{align}
where the constant $K_\Delta$ defined in Lemma~\ref{lem:boundsOnD} is numerically evaluated to  $K_\Delta \leq 2.32$ by setting the parameter $\alpha = 16.5$.

\paragraph{Step 2: bound $\left\| \bS \bP \bE(\btheta) \bS^{-1} \right\|_\infty$}
Using the block diagonal structure of the matrix $\bE(\btheta)$ defined in~\eqref{eq:E_definition}, we have that
\begin{align}\label{eq:E-calc-fixed-1}
	\bS \bP \bE(\btheta) \bS^{-1} = \begin{bmatrix}
		 \bm{0} & \frac{1}{\sqrt{-F_N^\pprime(0)}}\diag \left({\ba^\star}^{-1} \right) \bE_1(\btheta) \\
		 \frac{1}{\sqrt{-F_N^\pprime(0)}} \diag  \left( A^{-2} \ba^\star   \right)^\sH \bE_1(\btheta) & - \frac{1}{F_N^\pprime(0)}  A^{-2} \bE_2(\btheta)
	\end{bmatrix}.
\end{align}
Therefore, it follows
\begin{align}\label{eq:expendNormE_fix}
	\left\| \bS \bP \bE(\btheta) \bS^{-1} \right\|_\infty &\leq \frac{1}{\sqrt{-F_N^\pprime(0)}} \max_j \left\{ \max\left\{ \frac{1}{\left\vert a^\star_j\right\vert }, \frac{\left\vert a^\star_j \right\vert}{A^2} \right\} \left\vert \left\langle \delta^{\prime}_{\tau_j}, F_N \ast \left(\mu(\btheta) - \mu(\btheta^\star) \right) \right\rangle \right\vert \right\} \nonumber \\
	& \qquad - \frac{1}{F_N^\pprime(0)} \max_{j} \left\{ \frac{\left\vert a_j \right\vert}{ A^2} \left\vert \left\langle \delta^{\pprime}_{\tau_j}, F_N \ast \left(\mu(\btheta) - \mu(\btheta^\star) \right)  \right\rangle \right\vert   \right\} \nonumber \\
	&\leq  (a_{\min}^\star)^{-1} \left(1 + \left\Vert \bS(\btheta_k - \btheta^\star) \right\Vert_\infty \right) \Bigg( \frac{1}{\sqrt{-F_N^\pprime(0)}}  \max_j \left\{ \left\vert \left\langle \delta^{\prime}_{\tau_j}, F_N \ast \left( \mu(\btheta) - \mu \left(\btheta^\star \right) \right) \right\rangle \right\vert \right\} \nonumber \\
	& \ \qquad - \frac{1}{F_N^\pprime(0)} \max_{j} \left\{  \left\vert \left\langle \delta^{\pprime}_{\tau_j}, F_N \ast \left(\mu(\btheta) - \mu(\btheta^\star) \right)  \right\rangle \right\vert   \right\} \Bigg), \end{align}
where we used the bounds~\eqref{eq:A-S-bounds-1} in the second line.
From~\eqref{eq:expendNormE_fix}, it can be seen that bounding the quantity of interest amounts to controlling  $\left\vert \left\langle \delta^{\prime}_{\tau_j}, F_N \ast \left(\mu(\btheta) - \mu(\btheta^\star) \right)  \right\rangle \right\vert $ and $\left\vert    \left\langle \delta^{\pprime}_{\tau_j}, F_N \ast \left(\mu(\btheta) - \mu(\btheta^\star) \right)  \right\rangle \right\vert$, which can be achieved by applying Lemma~\ref{lem:boundOnEDev}. Substituting the expression provided by Lemma~\ref{lem:boundOnEDev} into~\eqref{eq:expendNormE_fix} leads to
\begin{multline}\label{eq:expendNormE_fix-2}
	\norm{\bS \bP \bE(\btheta) \bS^{-1}}_\infty \leq \Biggl( \left( C_1 \frac{(n+1)}{\sqrt{-F_N^\pprime(0)}} + C_2 \frac{\left( n+1 \right)^2}{-F_N^\pprime(0)} \right) \frac{\norm{\ba - \ba^\star}_\infty}{\norm{\ba^\star}_\infty} \\
	\qquad \qquad\qquad \qquad  + \left( C_2 \frac{\left(n+1\right)^2}{-F_N^\pprime(0)} + C_3 \frac{\left( n+1 \right)^3}{\left(-F_N^\pprime(0)\right)^{3/2}} \right) \sqrt{-F_N^\pprime(0)}\norm{\btau_k - \btau^\star}_\infty \Biggl) \\
	\cdot \frac{\norm{\ba^\star}_\infty}{a_{\min}^\star}\left( (n+1) \Delta(\btau) \right)^{-2} \left( 1 + \left\Vert \bS(\btheta_k - \btheta^\star) \right\Vert_\infty \right).
\end{multline}
Evaluating the constants $C_1 \leq 2.75$, $C_2 \leq 19.08$ and $C_3 \leq 48.74$ defined in~\eqref{eq:constant_values} with parameters $\alpha = 16.5$ and $\beta = \frac{\alpha}{4}=4.125$, altogether with the inequality $\frac{\left(n+1\right)^2}{-F_N^\pprime(0)}\leq \frac{27}{16 \pi^2}$ under the assumption $n \geq 2$ yields
\begin{multline}\label{eq:expendNormE_fix-3}
	\norm{\bS \bP \bE(\btheta) \bS^{-1}}_\infty \leq \left( K_a \frac{\norm{\ba - \ba^\star}_\infty}{\norm{\ba^\star}_\infty}
	 + K_\tau \sqrt{-F_N^\pprime(0)} \norm{\btau_k - \btau^\star}_\infty \right) \\
	 \cdot \frac{\norm{\ba^\star}_\infty}{a_{\min}^\star}\left( (n+1) \Delta(\btau) \right)^{-2} \left(1 + \left\Vert \bS(\btheta_k - \btheta^\star) \right\Vert_\infty  \right),
\end{multline}
where the constants $K_a$ and $K_\tau$ are given by
\begin{subequations}
	\begin{align}
		K_a & = C_1 \sqrt{\frac{27}{16\pi^2}} + C_2 \frac{27}{16\pi^2} \leq 4.40, \\
		K_\tau & =  C_2 \frac{27}{16\pi^2} + C_3 \left(\frac{27}{16\pi^2}\right)^{3/2}\leq 6.71.
	\end{align}
	\end{subequations}

\paragraph{Step 3: combine the bounds}
The scaled Hessian matrix $\bS \bP \bH(\btheta) \bS^{-1}$ can be controlled over $\cS_k$ by the triangle inequality as follows
\begin{align}
	\norm{\bS \bP \bH(\btheta) \bS^{-1} - \bI}_\infty & \leq \norm{\bS \bP \left( \bG(\btheta)  + \bE(\btheta) \right) \bS^{-1} - \bI}_\infty \nonumber\\
	& \leq \norm{\bS \bP  \bG(\btheta)   \bS^{-1} - \bI}_\infty + \norm{\bS \bP  \bE(\btheta) \bS^{-1}}_\infty.
\end{align}
Furthermore, we note that the inequality $\left\vert \Delta(\btau) - \Delta(\btau^\star) \right\vert \leq 2 \norm{\btau - \btau^\star}_\infty$ holds for every $\btau,\btau^\star \subset \bbR$. This implies, with the assumption $\norm{\btau - \btau^\star}_\infty \leq \frac{1}{4} \Delta(\btau^\star)$, that $\Delta(\btau) \geq \frac{1}{2} \Delta(\btau^\star)$. Substituting the bounds given in~\eqref{eq:bG_bound_fix} and~\eqref{eq:expendNormE_fix-3} yields
\begin{align}
	&  \norm{\bS \bP \bH(\btheta) \bS^{-1} - \bI}_\infty  \nonumber \\
	& \leq 1- \frac{{a_{\min}^\star}^2}{A^2}\left(1 - \norm{ \bS(\btheta_k - \btheta^\star)}_\infty \right)^2  \nonumber\\
	&	  \quad +\left( K_\Delta  + K_a \frac{\norm{\ba - \ba^\star}_\infty}{\norm{\ba^\star}_\infty}
	+ K_\tau \sqrt{-F_N^\pprime(0)} \norm{\btau_k - \btau^\star}_\infty \right)    \frac{\norm{\ba^\star}_\infty}{a_{\min}^\star} \left( (n+1) \Delta(\btau)\right)^{-2} \left( 1 + \norm{\bS \left(\btheta_k - \btheta^\star \right)}_\infty \right)^2 \nonumber \\
	& \leq 1 - \frac{{a_{\min}^\star}^2}{A^2}\left(1- \norm{ \bS(\btheta_k - \btheta^\star)}_\infty \right)^2  \nonumber\\
	&	  \quad +\left( K_\Delta  + \left(K_a + K_\tau \right) \norm{\bS \left( \btheta_k - \btheta^\star \right)}_\infty \right)  \frac{\norm{\ba^\star}_\infty}{a_{\min}^\star} \left( (n+1) \Delta(\btau)\right)^{-2} \left( 1 + \norm{\bS \left(\btheta_k - \btheta^\star \right)}_\infty \right)^2 \nonumber \\
	& \leq 1 - \frac{{a_{\min}^\star}^2}{A^2}\left(1 - \norm{ \bS(\btheta_k - \btheta^\star)}_\infty \right)^2  \nonumber\\
	&	  \quad +\left( K_\Delta  + \left(K_a + K_\tau \right) \norm{\bS \left( \btheta_k - \btheta^\star \right)}_\infty \right)  \frac{\norm{\ba^\star}_\infty}{a_{\min}^\star} 4\left( (n+1) \Delta(\btau^\star)\right)^{-2} \left( 1 + \norm{\bS \left(\btheta_k - \btheta^\star \right)}_\infty \right)^2 \nonumber \\
	& \leq 1 - \frac{{a_{\min}^\star}^2}{A^2}\left(1 - \norm{ \bS(\btheta_k - \btheta^\star)}_\infty \right)^2  \nonumber\\
	&	  \quad +\left( 4 K_\Delta  + K_\theta \norm{\bS \left( \btheta_k - \btheta^\star \right)}_\infty \right)  \frac{\norm{\ba^\star}_\infty}{a_{\min}^\star} \left( (n+1) \Delta(\btau^\star)\right)^{-2} \left( 1 + \norm{\bS \left(\btheta_k - \btheta^\star \right)}_\infty \right)^2,
\end{align}
where we defined in the last line the constant $K_\theta$ as
\begin{equation}\label{eq:K_theta}
	K_\theta = 4 \left(K_a + K_\tau\right) \leq 44.42 .
\end{equation}
This concludes the proof of the theorem. \qed

\subsection{Proof of Theorem \ref{theo:uniformBounds}}\label{subsec:proof_uniformBounds_adaptive}

We proceed analogously to the proof of Theorem~\ref{theo:uniformBounds_fixed} presented in Appendix~\ref{subsec:proof_uniformBounds_fixed}. We start from the expansion~\eqref{eq:H_decomposition} of the Hessian matrix to get
\begin{equation}
	\bS \bP_k \bH(\btheta) \bS^{-1}    - \bI  = \bS \bP_k \bG(\btheta) \bS^{-1} - \bI   +  \bS \bP_k \bE(\btheta) \bS^{-1},
\end{equation}
and proceed to bound $\left\|\bS \bP_k \bG(\btheta) \bS^{-1} - \bI\right\|_{\infty}$ and $\left\| \bS \bP_k \bE(\btheta) \bS^{-1} \right\|_\infty$ individually before recombining them via the triangle inequality.

\paragraph{Step 1: bound $\left\| \bS \bP_k \bG(\btheta) \bS^{-1}    - \bI  \right\|_\infty$}
From~\eqref{eq:S_definition} and~\eqref{eq:G_definition}, we have that
\begin{align}
	  \bS \bP_k \bG(\btheta) \bS^{-1}    - \bI
	& = \bS \bP_k \diag \left(\begin{bmatrix}
		\bm{1}_r \\
		\sqrt{-F_N^\pprime(0)} \ba
		\end{bmatrix} \right)^{\sH} 	\bD(\btau)
		\diag \left(\begin{bmatrix}
			\bm{1}_r \\
			\sqrt{-F_N^\pprime(0)} \ba^\sH
			\end{bmatrix} \right)  \bS^{-1} - \bI \nonumber\\
	& = \diag \left(\begin{bmatrix}
			{\ba^\star}^{-1} \\
			{\left\vert \ba_k \right\vert^{-2}} \odot \ba
			\end{bmatrix} \right)^\sH 	\bD(\btau)
			\diag \left(\begin{bmatrix}
			\ba^\star \\
			\ba
			\end{bmatrix} \right) - \bI \nonumber\\
			& = \diag \left(\begin{bmatrix}
				{\ba^\star}^{-1} \\
			 	{\left\vert \ba_k \right\vert^{-2}} \odot \ba
				\end{bmatrix} \right)^\sH 	\left( \bD(\btau) - \bI \right)
				\diag \left(\begin{bmatrix}
				\ba^\star \\
				\ba
				\end{bmatrix} \right) + \diag \left(\begin{bmatrix}
					\bm{1}_r \\
					{\left\vert \ba_k \right\vert^{-2}} \odot \left\vert \ba \right\vert^2
					\end{bmatrix} \right) - \bI.
\end{align}
This immediately yields from the triangle inequality
\begin{align}\label{eq:bG_bound_before}
	\MoveEqLeft \norm{ \bS \bP_k \bG(\btheta) \bS^{-1}    - \bI}_\infty & \nonumber \\
	& \leq \norm{\diag \left(\begin{bmatrix}
		{\ba^\star}^{-1} \\
		{\left\vert \ba_k \right\vert^{-2}} \odot \ba
		\end{bmatrix} \right)^\sH 	\left( \bD(\btau) - \bI \right)
		\diag \left(\begin{bmatrix}
		\ba^\star \\
		\ba
		\end{bmatrix} \right)}_\infty + \norm{\diag \left(\begin{bmatrix}
			\bm{1}_r \\
			{\left\vert \ba_k \right\vert^{-2}} \odot \left\vert \ba \right\vert^2
			\end{bmatrix} \right) - \bI}_\infty \nonumber \\
	& \leq \norm{\diag \left(\begin{bmatrix}
		{\ba^\star}^{-1} \\
		{\left\vert \ba_k \right\vert^{-2}} \odot \ba
		\end{bmatrix} \right)^\sH}_\infty \norm{	  \bD(\btau) - \bI   }_\infty
		\norm{ \diag \left(\begin{bmatrix}
		\ba^\star \\
		\ba
		\end{bmatrix} \right)}_\infty + \norm{\diag \left(\begin{bmatrix}
			\bm{1}_r \\
			{\left\vert \ba_k \right\vert^{-2}} \odot \left\vert \ba \right\vert^2
			\end{bmatrix} \right) - \bI}_\infty \nonumber \\
	&\leq \max_j \left\{\frac{1}{\left\vert a_j^\star \right\vert}, \frac{ \left\vert a_{j}^\star \right\vert}{\left\vert a_{k,j} \right\vert^2} \right\} \max_j \left\{ \left\vert a_j^\star \right\vert , \left\vert a_j  \right\vert \right\} \norm{   \bD(\btau) - \bI  }_\infty + \max_j \left\{ \left| \frac{ \left\vert a_{j}^\star \right\vert^2}{\left\vert a_{k,j} \right\vert^2} -1   \right| \right\}
	\end{align}
	The three maxima in~\eqref{eq:bG_bound_before} can be controlled using the basic relation
	\begin{align}\label{eq:dynamic_err_cnt}
		\frac{\left\vert a_j^{\star} \right\vert}{\left\vert a_{k,j} \right\vert} & \leq \frac{ \left\vert a_{j}^\star \right\vert}{ \left\vert a_{j}^\star \right\vert - \left\vert a_{k,j} - a_j^\star \right\vert} \leq \frac{1}{1 - \left\Vert \bS(\btheta_k - \btheta^\star) \right\Vert}_\infty,
	\end{align}
	as follows
\begin{subequations}\label{eq:A-S-bounds-2}
	\begin{align}
		\max_j \left\{\frac{1}{\left\vert a_j^\star \right\vert}, \frac{ \left\vert a_{j}^\star \right\vert}{\left\vert a_{k,j} \right\vert^2} \right\} &\leq \max_j \left\{\frac{1}{\left\vert a_j^\star \right\vert}, \frac{ 1}{\left\vert a_j^\star \right\vert} \frac{1}{ \left( 1 - \norm{\bS (\btheta_k - \btheta^\star)}_\infty \right)^2 }   \right\} \leq \frac{1}{a_{\min}^\star} \frac{1}{\left(1 - \norm{\bS (\btheta_k - \btheta^\star)}_\infty \right)^2}  \\
		\max_j \left\{ \left\vert a_j^\star \right\vert , \left\vert a_j  \right\vert \right\} &\leq \max_j \left\{ \left\vert a_j^\star \right\vert , \left\vert a_j^\star \right\vert \left( 1 + \norm{\bS (\btheta_k - \btheta^\star)}_\infty \right)  \right\} \leq \norm{\ba^\star}_\infty \left( 1 + \norm{\bS (\btheta_k - \btheta^\star)}_\infty \right) \\
		\max_j \left\{ \left| \frac{ \left\vert a_{j}^\star \right\vert^2}{\left\vert a_{k,j} \right\vert^2} -1   \right| \right\} &\leq \max_j \left\{ 1- \frac{ \left\vert a_{j}^\star \right\vert^2}{\left\vert a_{k,j} \right\vert^2}, \frac{ \left\vert a_{j}^\star \right\vert^2}{\left\vert a_{k,j} \right\vert^2} -1  \right\} \nonumber \\
		& \leq \max_j \left\{1, \frac{ \left\vert a_{j}^\star \right\vert^2}{\left\vert a_{k,j} \right\vert^2} -1 \right\} \leq \frac{1}{\left( 1 - \norm{\bS (\btheta_k - \btheta^\star)}_\infty \right)^2} - 1.
	\end{align}
\end{subequations}
The bounds~\eqref{eq:A-S-bounds-2} and Lemma~\ref{lem:boundsOnD} conclude with~\eqref{eq:bG_bound_before} on
\begin{align}\label{eq:bG_bound}
	\MoveEqLeft \norm{ \bS \bP_k \bG(\btheta) \bS^{-1}    - \bI}_\infty & \nonumber \\
	&\leq \frac{\norm{\ba^\star}_\infty}{a_{\min}^\star}\frac{1 + \norm{\bS \left( \btheta_k - \btheta^\star \right)}_\infty}{\left(1 - \norm{ \bS(\btheta_k - \btheta^\star)}_\infty \right)^2} \norm{  \bD(\btau) - \bI   }_\infty + \frac{1}{\left(1 - \norm{ \bS(\btheta_k - \btheta^\star)}_\infty \right)^2} - 1  \nonumber\\
	&\leq \frac{\norm{\ba^\star}_\infty}{a_{\min}^\star}\frac{1 + \norm{\bS \left( \btheta_k - \btheta^\star \right)}_\infty}{\left(1 - \norm{ \bS(\btheta_k - \btheta^\star)}_\infty \right)^2} K_\Delta \left((n+1) \Delta(\btau)\right)^{-2} + \frac{1}{\left(1 - \norm{ \bS(\btheta_k - \btheta^\star)}_\infty \right)^2} - 1,
\end{align}
where the constant $K_\Delta$ defined in Lemma~\ref{lem:boundsOnD} is numerically evaluated to  $K_\Delta \leq 2.32$ by setting the parameter $\alpha = 4.7$.

\paragraph{Step 2: bound $\left\| \bS \bP_k \bE(\btheta) \bS^{-1} \right\|_\infty$}
Again, using the block diagonal structure of the matrix $\bE(\btheta)$ defined in~\eqref{eq:E_definition}, we similarly have that
\begin{align}
	\bS \bP_k \bE(\btheta) \bS^{-1} = \begin{bmatrix}
		 \bm{0} & \frac{1}{\sqrt{-F_N^\pprime(0)}}\diag \left({\ba^\star}^{-1} \right) \bE_1(\btheta) \\
		 \frac{1}{\sqrt{-F_N^\pprime(0)}} \diag \left( \left\vert \ba_k \right\vert^{-2} \odot \ba^\star   \right)^\sH \bE_1(\btheta) & - \frac{1}{F_N^\pprime(0)} \left\vert \ba_k \right\vert^{-2} \bE_2(\btheta)
	\end{bmatrix}.
\end{align}
Therefore, the quantity of interest can be bounded as follows
\begin{align}\label{eq:expendNormE}
	\left\| \bS \bP_k \bE(\btheta) \bS^{-1} \right\|_\infty &\leq \frac{1}{\sqrt{-F_N^\pprime(0)}} \max_j \left\{ \max\left\{ \frac{1}{\left\vert a^\star_j\right\vert }, \frac{\left\vert a^\star_j \right\vert}{\left\vert a_{j,k} \right\vert^2} \right\} \left\vert \left\langle \delta^{\prime}_{\tau_j}, F_N \ast \left(\mu(\btheta) - \mu(\btheta^\star) \right) \right\rangle \right\vert \right\} \nonumber \\
	& \qquad - \frac{1}{F_N^\pprime(0)} \max_{j} \left\{ \frac{\left\vert a_j \right\vert}{\left\vert a_{j,k} \right\vert^2} \left\vert \left\langle \delta^{\pprime}_{\tau_j}, F_N \ast \left(\mu(\btheta) - \mu(\btheta^\star) \right)  \right\rangle \right\vert   \right\} \nonumber \\
	&\leq  \frac{(a_{\min}^\star)^{-1}}{\left(1 - \left\Vert \bS(\btheta_k - \btheta^\star) \right\Vert_\infty  \right)^2} \Bigg( \frac{1}{\sqrt{-F_N^\pprime(0)}}  \max_j \left\{ \left\vert \left\langle \delta^{\prime}_{\tau_j}, F_N \ast \left( \mu(\btheta) - \mu \left(\btheta^\star \right) \right) \right\rangle \right\vert \right\} \nonumber \\
	& \qquad \qquad - \frac{1}{F_N^\pprime(0)} \max_{j} \left\{  \left\vert \left\langle \delta^{\pprime}_{\tau_j}, F_N \ast \left(\mu(\btheta) - \mu(\btheta^\star) \right)  \right\rangle \right\vert   \right\} \Bigg), \end{align}
where we used the inequalities~\eqref{eq:A-S-bounds-2} on the second line.
Substituting the expression provided by Lemma~\ref{lem:boundOnEDev} into~\eqref{eq:expendNormE} leads to
\begin{multline}\label{eq:expendNormE-2}
	\norm{\bS \bP_k \bE(\btheta) \bS^{-1}}_\infty \leq  \Biggl( \left( C_1 \frac{n+1}{\sqrt{-F_N^\pprime(0)}}
	+ C_2 \frac{\left( n+1 \right)^2}{-F_N^\pprime(0)} \right) \frac{\norm{\ba - \ba^\star}_\infty}{\norm{\ba^\star}_\infty} \\
	\qquad \qquad \qquad + \left( C_2 \frac{\left(n+1\right)^2}{-F_N^\pprime(0)} + C_3 \frac{\left( n+1 \right)^3}{\left(-F_N^\pprime(0)\right)^{3/2}} \right) \sqrt{-F_N^\pprime(0)}\norm{\btau_k - \btau^\star}_\infty \Biggl) \\
	\cdot \frac{\norm{\ba^\star}_\infty}{a_{\min}^\star} \frac{\left( (n+1) \Delta(\btau) \right)^{-2}}{\left(1 - \left\Vert \bS(\btheta_k - \btheta^\star) \right\Vert_\infty  \right)^2}.
\end{multline}
Evaluating the constants $C_1 \leq 3.24$, $C_2 \leq 22.90$ and $C_3 \leq 105.55$ defined in~\eqref{eq:constant_values} with parameters $\alpha = 4.7$ and $\beta = \frac{\alpha}{4} = 1.2$, altogether with the inequality $\frac{\left(n+1\right)^2}{-F_N^\pprime(0)}\leq \frac{27}{16 \pi^2}$ under the assumption $n \geq 2$ yields
\begin{multline}\label{eq:expendNormE-3}
	\norm{\bS \bP_k \bE(\btheta) \bS^{-1}}_\infty
	\leq \left( K_a \frac{\norm{\ba - \ba^\star}_\infty}{\norm{\ba^\star}_\infty}
	 + K_\tau \sqrt{-F_N^\pprime(0)} \norm{\btau_k - \btau^\star}_\infty \right)   \frac{\norm{\ba^\star}_\infty}{a_{\min}^\star} \frac{ \left( (n+1) \Delta(\btau) \right)^{-2}}{\left(1 - \left\Vert \bS(\btheta_k - \btheta^\star) \right\Vert_\infty  \right)^2},
\end{multline}
where the constants $K_a$ and $K_\tau$ are given by
\begin{subequations}
	\begin{align}
		K_a & = C_1 \sqrt{\frac{27}{16\pi^2}} + C_2 \frac{27}{16\pi^2} \leq 5.26, \\
		K_\tau & =  C_2 \frac{27}{16\pi^2} + C_3 \left(\frac{27}{16\pi^2}\right)^{3/2}\leq 11.38.
	\end{align}
	\end{subequations}

\paragraph{Step 3: combine the bounds}
The scaled Hessian matrix $\bS \bP_k \bH(\btheta) \bS^{-1}$ can be controlled over $\cS_k$ by the triangle inequality as follows
\begin{align}
	\norm{\bS \bP_k \bH(\btheta) \bS^{-1} - \bI}_\infty & \leq \norm{\bS \bP_k \left( \bG(\btheta)  + \bE(\btheta) \right) \bS^{-1} - \bI}_\infty \nonumber\\
	& \leq \norm{\bS \bP_k   \bG(\btheta)   \bS^{-1} - \bI}_\infty + \norm{\bS \bP_k  \bE(\btheta) \bS^{-1}}_\infty.
\end{align}
Finally, the inequality $\left\vert \Delta(\btau) - \Delta(\btau^\star) \right\vert \leq 2 \norm{\btau - \btau^\star}_\infty$ holds for every $\btau,\btau^\star \subset \bbR$. This implies, with the assumption $\norm{\btau - \btau^\star}_\infty \leq \frac{1}{4} \Delta(\btau^\star)$, that $\Delta(\btau) \geq \frac{1}{2} \Delta(\btau^\star)$. Substituting the bounds given in~\eqref{eq:bG_bound} and~\eqref{eq:expendNormE-3} yields
\begin{align}
& \norm{\bS \bP_k \bH(\btheta) \bS^{-1} - \bI}_\infty   \nonumber \\
	& \leq \frac{1}{\left(1 - \norm{ \bS(\btheta_k - \btheta^\star)}_\infty \right)^2} - 1 \nonumber\\
	&	+\left( K_\Delta \left(1 + \norm{\bS \left(\btheta_k - \btheta^\star \right)}_\infty \right) + K_a \frac{\norm{\ba - \ba^\star}_\infty}{\norm{\ba^\star}_\infty}
	+ K_\tau \sqrt{-F_N^\pprime(0)} \norm{\btau_k - \btau^\star}_\infty \right)  \frac{\norm{\ba^\star}_\infty}{a_{\min}^\star} \frac{\left( (n+1) \Delta(\btau)\right)^{-2}}{\left( 1 - \norm{\bS \left(\btheta_k - \btheta^\star \right)}_\infty \right)^2} \nonumber \\
	& \leq  \frac{1}{\left(1 - \norm{ \bS(\btheta_k - \btheta^\star)}_\infty \right)^2} - 1 \nonumber \\
	&   \qquad+\left( K_\Delta \left(1 + \norm{\bS \left(\btheta_k - \btheta^\star \right)}_\infty \right) + \left(K_a + K_\tau \right) \norm{\bS \left(\btheta_k - \btheta^\star\right)}_\infty \right)    \frac{\norm{\ba^\star}_\infty}{a_{\min}^\star} \frac{\left( (n+1) \Delta(\btau)\right)^{-2}}{\left( 1 - \norm{\bS \left(\btheta_k - \btheta^\star \right)}_\infty \right)^2} \nonumber\\
	& \leq   \frac{1}{\left(1 - \norm{ \bS(\btheta_k - \btheta^\star)}_\infty \right)^2} - 1 \nonumber \\
	& \qquad + \left(K_\Delta + \left(K_\Delta + K_a + K_\tau \right) \norm{\bS \left(\btheta_k - \btheta^\star \right)}_\infty  \right)    \frac{\norm{\ba^\star}_\infty}{a_{\min}^\star}  \frac{\left( (n+1) \Delta(\btau)\right)^{-2}}{\left( 1 - \norm{\bS \left(\btheta_k - \btheta^\star \right)}_\infty \right)^2} \nonumber \\
	& \leq   \frac{1}{\left(1 - \norm{ \bS(\btheta_k - \btheta^\star)}_\infty \right)^2} - 1 \nonumber \\
	& \qquad + \left( K_\Delta +  \left(K_\Delta + K_a + K_\tau \right) \norm{\bS \left(\btheta_k - \btheta^\star \right)}_\infty  \right)    \frac{\norm{\ba^\star}_\infty}{a_{\min}^\star}  \frac{4 \left((n+1) \Delta(\btau^\star)\right)^{-2}}{\left( 1 - \norm{\bS \left(\btheta_k - \btheta^\star \right)}_\infty \right)^2} \nonumber \\
	& \leq \frac{1}{\left(1 - \norm{ \bS(\btheta_k - \btheta^\star)}_\infty \right)^2} - 1 + \left(4 K_\Delta + K_\theta \norm{\bS \left(\btheta_k - \btheta^\star \right)}_\infty  \right)    \frac{\norm{\ba^\star}_\infty}{a_{\min}^\star} \frac{\left( (n+1) \Delta(\btau^\star) \right)^{-2}}{\left(1 - \left\Vert \bS(\btheta_k - \btheta^\star) \right\Vert_\infty  \right)^2},
\end{align}
where defined the constant $K_\theta$ as
\begin{equation}\label{eq:K_theta_adap}
	K_\theta = 4 \left( K_\Delta + K_a + K_\tau\right) \leq 75.80.
\end{equation}
This concludes the proof of the theorem. \qed

\subsection{Proof of Lemma \ref{lem:boundsOnD}}\label{subsec:proof_boundOnD}

Leveraging the block structure~\eqref{eq:D_definition} of the matrix $\bD(\btau)$, it boils down to controlling
\begin{equation}\label{eq:boundD-2}
	\left\Vert \bD(\btau) - \bI \right\Vert_\infty \leq \max \left\{ \left\Vert \bD_0(\btau) - \bI \right\Vert_\infty + \left\Vert \bD_1(\btau) \right\Vert_\infty, \left\Vert \bD_2(\btau) - \bI \right\Vert_\infty + \left\Vert \bD_1(\btau) \right\Vert_\infty  \right\},
\end{equation}
which can be accomplished with the aid of Lemma \ref{lem:BoundsFejerKernel}. Specifically, recalling the expressions in~\eqref{eq:D_expressions}, applying Lemma \ref{lem:BoundsFejerKernel} with parameters $\alpha$ and $\beta=0$, and noticing that the inequality $\frac{\left( n+1 \right)^2}{-F_N^\pprime(0)} \leq \frac{27}{16\pi^2}$ holds whenever $n \geq 2$, the quantities of interest in~\eqref{eq:boundD-2} can be controlled as follows
\begin{subequations}\label{eq:boundD-blocks}
	\begin{align}
		\left\Vert \bD_0(\btau) - \bI \right\Vert_\infty & = \max_i \sum_{j\neq i} \left\vert F_N(\tau_j - \tau_i) \right\vert \leq C_0 \left((n+1)\Delta(\btau)\right)^{-2}, \\
		\left\Vert \bD_1(\btau) \right\Vert_\infty & = \frac{1}{\sqrt{- F_N^{\pprime}(0)}} \max_i \sum_{j} \left\vert F_N^\prime(\tau_j - \tau_i) \right\vert \leq C_1 \frac{n+1}{\sqrt{-F_N^{\pprime}(0)}} \left( (n+1) \Delta(\btau)\right)^{-2} \nonumber \\
		&\leq  \frac{3\sqrt{3}}{4\pi} C_1 \left( (n+1) \Delta(\btau)\right)^{-2}, \\
		\left\Vert \bD_2(\btau) - \bI \right\Vert_\infty &= \frac{1}{- F_N^{\pprime}(0)} \max_i  \sum_{j\neq i} \left\vert F_N^\pprime(\tau_j - \tau_i) \right\vert \leq C_2 \frac{\left(n+1\right)^2}{-F_N^\pprime(0)} \left( (n+1) \Delta(\btau)\right)^{-2} \nonumber \\
		&\leq  \frac{27}{16 \pi^2} C_2  \left( (n+1) \Delta(\btau)\right)^{-2}.
	\end{align}
\end{subequations}
Further substituting~\eqref{eq:boundD-blocks} into~\eqref{eq:boundD-2} leads to
\begin{align}
	\left\Vert \bD(\btau) - \bI \right\Vert_\infty & \leq   \max \left\{ C_0 + \frac{3\sqrt{3}}{4\pi} C_1, \frac{3\sqrt{3}}{4\pi} C_1 + \frac{27}{16 \pi^2} C_2 \right\}
	\left( (n+1) \Delta(\btau)\right)^{-2}
\end{align}
which leads to the desired statement. \qed

\subsection{Proof of Lemma \ref{lem:boundOnEDev}}\label{subsec:proof_boundOnE}
For the first inequality, following the definition, we have that
\begin{align}\label{eq:lem4-1}
	\left\vert \left\langle \Phi(\delta^{\prime}_{\tau_j}), \Phi \left(\mu(\btheta) - \mu(\btheta_\star) \right) \right\rangle \right\vert & = \left\vert \left\langle \delta^{\prime}_{\tau_j}, \Phi^\ast \left( \Phi \left(\mu(\btheta) - \mu(\btheta_\star) \right) \right) \right\rangle \right\vert \nonumber \\
	& = \left\vert \sum_{\ell=1}^r a_\ell F_N^{\prime}(\tau_j - \tau_\ell) - \sum_{\ell=1}^{r^\star} a_\ell^\star F_N^{\prime}(\tau_j - \tau_\ell^\star) \right\vert \nonumber \\
	& =  \left\vert \sum_{\ell=1}^r (a_\ell - a_\ell^\star) F_N^{\prime} (\tau_j - \tau_\ell) + \sum_{\ell=1}^r a_\ell^\star \left( F_N^\prime (\tau_j - \tau_\ell) - F_N^\prime (\tau_j - \tau^\star_\ell)  \right)\right\vert \nonumber \\
	& \leq \left\vert \sum_{\ell=1}^r (a_\ell - a_\ell^\star) F_N^{\prime} (\tau_j - \tau_\ell) \right\vert + \left\vert \sum_{\ell=1}^r a_\ell^\star \left( F_N^\prime (\tau_j - \tau_\ell) - F_N^\prime (\tau_j - \tau^\star_\ell)  \right) \right\vert,
\end{align}
where the last line used the triangle inequality. To proceed, we control the two terms separately.
Using Hölder's inequality and Lemma~\ref{lem:BoundsFejerKernel}, we obtain
\begin{align}
	\left\vert \sum_{\ell=1}^r (a_\ell - a_\ell^\star) F_N^{\prime} (\tau_j - \tau_\ell) \right\vert & \leq \left\Vert \ba - \ba^\star \right\Vert_\infty  \sum_{\ell=1}^r \left\vert F_N^\prime (\tau_j - \tau_\ell) \right\vert \nonumber \\
	& \leq C_1 (n + 1) \left\Vert \ba - \ba^\star \right\Vert_\infty \left((n+1)\Delta(\btau)\right)^{-2}.
\end{align}
Next, the second term can be bounded by applying H{\"o}lder's inequality, the mean-value theorem and Lemma~\ref{lem:BoundsFejerKernel} with parameter $\beta \geq (n+1)\norm{\btau - \btau^\star}_\infty$ as follows
\begin{align}
	\left\vert \sum_{\ell=1}^r a_\ell^\star \left( F_N^\prime (\tau_j - \tau_\ell) - F_N^\prime (\tau_j - \tau^\star_\ell)  \right) \right\vert \leq{}& \left\Vert \ba^\star \right\Vert_\infty \sum_{\ell=1}^r  \left\vert F_N^\prime (\tau_j - \tau_\ell) - F_N^\prime (\tau_j - \tau^\star_\ell) \right\vert \nonumber  \\
	\leq{}& \left\Vert \ba^\star \right\Vert_\infty \left\Vert \btau - \btau^\star \right\Vert_{\infty} \sum_{\ell=1}^r  \sup_{ \left\vert u_\ell \right\vert \leq \left\Vert \btau - \btau^\star \right\Vert_{\infty}}\left\vert F_N^\pprime (\tau_j - \tau_\ell + u_\ell) \right\vert  \nonumber \\
	\leq{}& \left\Vert \ba^\star \right\Vert_\infty \left\Vert \btau - \btau^\star \right\Vert_{\infty} C_2 {(n+1)}^2 \left( (n+1) \Delta(\btau)\right)^{-2}.
\end{align}
Plugging the previous two bounds into the inequality~\eqref{eq:lem4-1} reduces to
\begin{multline}
	\left\vert \left\langle \Phi(\delta^{\prime}_{\tau_j}), \Phi \left(\mu(\btheta) - \mu(\btheta_\star) \right) \right\rangle \right\vert \\
	 \leq \left(C_1 \left\Vert \ba - \ba^\star \right\Vert_\infty + C_2  \left\Vert \ba^\star \right\Vert_\infty (n+1)\left\Vert \btau - \btau^\star \right\Vert_\infty \right) (n+1)\left( (n+1) \Delta(\btau)\right)^{-2} .
\end{multline}
Moreover, we may show that through analogous reasoning that
\begin{multline}
	\left\vert \left\langle \Phi(\delta^{\pprime}_{\tau_j}), \Phi \left(\mu(\btheta) - \mu(\btheta_\star) \right) \right\rangle \right\vert \\
	\leq \left(C_2 \left\Vert \ba - \ba^\star \right\Vert_\infty + C_3  \left\Vert \ba^\star \right\Vert_\infty (n+1)\left\Vert \btau - \btau^\star \right\Vert_\infty \right) {(n+1)}^2 \left( (n+1) \Delta(\btau)\right)^{-2},
\end{multline}
which concludes the proof. \qed

\renewcommand*{\bibfont}{\footnotesize}
\printbibliography

@article{chizat2018global,
  title={On the global convergence of gradient descent for over-parameterized models using optimal transport},
  author={Chizat, Lenaic and Bach, Francis},
  journal={Advances in neural information processing systems},
  volume={31},
  year={2018}
}

@article{malioutov2005sparse,
  title={A sparse signal reconstruction perspective for source localization with sensor arrays},
  author={Malioutov, Dmitry and Cetin, M{\"u}jdat and Willsky, Alan S},
  journal={IEEE transactions on signal processing},
  volume={53},
  number={8},
  pages={3010--3022},
  year={2005},
  publisher={IEEE}
}

\end{document}